\documentclass[12pt]{article}
\usepackage{epsfig}
\usepackage{graphicx}   
\usepackage{amsmath}
\usepackage{amssymb}
\newcommand{\nn}{\nonumber}

\newcommand{\benu}{\begin{enumerate}}
\newcommand{\eenu}{\end{enumerate}}
\newcommand{\norsl}{\normalsize\sl}
\newcommand{\norsc}{\normalsize\sc}
\newcommand{\noi}{\noindent}

\def\Fv{\mbox{\boldmath $F$}}

\def\Lv{\mbox{\boldmath $L$}}

\def\Nv{\mbox{\boldmath $N$}}
\def\Xv{\mbox{\boldmath $X$}}

\def\kv{\mbox{\boldmath $k$}}

\def\ev{\mbox{\boldmath $e$}}

\def\Wv{\mbox{\boldmath $W$}}
\def\uv{\mbox{\boldmath $u$}}
\def\vv{\mbox{\boldmath $v$}}
\def\xv{\mbox{\boldmath $x$}}
\def\yv{\mbox{\boldmath $y$}}
\def\zv{\mbox{\boldmath $z$}}
\def\omegav{\mbox{\boldmath $\omega$}}
\def\Omegav{\mbox{\boldmath $\Omega$}}
\def\zerov{\mbox{\boldmath $0$}}
\def\etav{\mbox{\boldmath $\eta$}}
\def\sv{\mbox{\boldmath $s$}}

\makeatletter
         
          \@addtoreset{equation}{section}
\makeatother

\begin{document}

\begin{titlepage}

\title{Motion of the Tippe Top \\
Gyroscopic Balance Condition and Stability
}
\author{
\norsc Takahiro UEDA\thanks{e-mail address: t-ueda@phys.ynu.ac.jp}~,
     Ken SASAKI\thanks{e-mail address: sasaki@phys.ynu.ac.jp}~ and
           Shinsuke WATANABE\thanks{e-mail address: 
wtnb@ynu.ac.jp} \\
\norsl  Dept. of Physics,  Faculty of Engineering, Yokohama National
University \\
\norsl  Yokohama 240-8501, JAPAN \\
}

\date{}
\maketitle

\begin{abstract}
{\normalsize
We reexamine a very classical problem, the spinning behavior of the tippe top on a horizontal table. 
The analysis is made for an eccentric sphere version of the tippe top, assuming 
a modified Coulomb law for the sliding friction,  which is a continuous function of the slip 
velocity $\vv_P$  at the point of contact and vanishes  at $\vv_P\!=\!\zerov$. 
We study the relevance 
of the gyroscopic balance condition (GBC), which was discovered to hold for a rapidly 
spinning hard-boiled egg by Moffatt and Shimomura, to the inversion phenomenon of the tippe top. 
We introduce a variable $\xi$ so that $\xi\!=\!0$ corresponds to the GBC and analyze the behavior of $\xi$.
Contrary to the case of the spinning egg, the GBC for the tippe top is not fulfilled initially. 
But we find from simulation that 
for those tippe tops which will turn over, the GBC will soon  be  satisfied approximately. 
It is shown that the GBC and the geometry  lead to 
the classification of tippe tops into three groups: 
The tippe tops of  Group I never flip over however large a spin they are given. 
Those of Group II show a complete inversion  and the tippe tops of Group III tend to turn over up to 
a certain inclination angle $\theta_f$ such that $\theta_f\!<\!\pi$, 
when they are spun sufficiently rapidly. 
There exist three steady states for the spinning motion of the tippe top. 
Giving a new criterion for  stability, we examine the stability of these states 
in terms of the initial spin velocity $n_0$.
And we obtain a critical value $n_c$ of the initial spin  which is required for the tippe top of Group II 
to flip over  up to the completely inverted position. 
}
\end{abstract}
\begin{picture}(5,2)(-270,-520)
\put(2.3,35){YNU-HEPTh-05-102}
\put(2.3,20){July 2005}
\end{picture}

\thispagestyle{empty}
\end{titlepage}
\setcounter{page}{1}
\baselineskip 18pt
\section{Introduction}
\smallskip
Spinning objects have historically been interesting subjects to study. 
The spin reversal of the rattleback~\cite{GH} (also called a celt
or wobblestone) and the behavior of the tippe top are typical examples. 
In the latter case, when a truncated sphere with a cylindrical stem, a so-called 
`tippe top',  is spun sufficiently rapidly on a table with its stem up, it will 
flip over and rotate on its stem. This inversion phenomenon has fascinated 
physicists and has been studied for over a century~\cite{Braams,Hugenholtz,Cohen,
Or,Leutwyler,EbenfeldScheck,GN,BMR}.

In the present paper we revisit and study this very classical problem from a different perspective. 
Recently the riddle of spinning eggs has been resolved by 
Moffatt and Shimomura [MS] \cite{MS}.  
They discovered that if an axisymmetric 
body, such as a hard-boiled egg, is spun sufficiently rapidly, a  `{\it gyroscopic balance}' condition 
(GBC) holds  and that under this condition the governing equations of the system are much simplified. 
In particular,  they derived a first-order ordinary differential equation (ODE) for $\theta$, 
the angle between the axis of symmetry and the vertical axis,  
and showed for the case of a prolate spheroid that the axis of symmetry 
indeed rises from the horizontal to the vertical. Then 
the spinning behavior of egg-shaped axisymmetric bodies, whose
cross sections are described by several models of oval curves, was
studied under the  GBC by one of the present authors~\cite{Sasaki}. 

The tippe top is also an 
axisymmetric body and shows the similar behavior as the spinning egg. 
Then one may ask: does the GBC also hold for  the tippe top?  
If so, how is it related to the inversion phenomenon of the tippe top? 
In the first half of this paper we analyze the spinning motion of the tippe top in terms of the GBC. 
Actually the GBC is not satisfied initially for the tippe top, contrary to the case of the spinning egg. 
The difference comes from how we start to spin the object: we spin the tippe top with its 
stem up, in other words, with its symmetry axis vertical while the egg is spun 
with its symmetry axis horizontal. 
In this paper we perform our analysis taking an eccentric sphere version of the tippe top 
instead of a commercially available one, a truncated sphere with a cylindrical stem. 
In order to examine the GBC of the tippe top more closely, we introduce a variable $\xi$ so that 
$\xi\!=\!0$ corresponds to the GBC, and study the behavior of $\xi$. 
Numerical analysis shows that 
for the tippe tops which will turn over, 
the variable $\xi$, starting from a large positive value $\xi_0$, 
soon takes negative values and fluctuates around a negative but small value $\xi_m$  
such that $|\xi_m/\xi_0|\approx 0$. Thus for these tippe tops,  the GBC, which is not satisfied initially, 
will soon  be  realized but approximately.   
On the other hand,  in the case of the tippe tops which 
will not turn over,  $\xi$ remains positive around $\xi_0$ or changes from  positive $\xi_0$ to negative 
values and then 
back to  positive values close to $\xi_0$ again. We find that the behavior of  $\xi$ is closely related to 
the inversion phenomenon of the tippe top. Once $\xi$ fluctuates around the value $\xi_{\rm m}$, 
the system becomes unstable and starts to turn over.

Under the GBC the governing equations for the tippe top are much 
simplified  and we obtain a first-order ODE for $\theta$,
which has the same form as the one derived by MS for the spinning egg. 
Then, this equation for $\theta$ and the geometry lead to 
the classification of tippe tops into {\it three} groups, depending on the values of $\frac{A}{C}$ 
and $\frac{a}{R}$, where $A$ and $C$ are two principal moments of inertia, 
and $a$ is the distance from the center of sphere to the center of mass 
and $R$ is the radius  of sphere. The tippe tops of  Group I never flip over however large a spin 
they are given.  Those of Group II show a complete inversion 
and the tippe tops of Group III tend to turn over up to 
a certain inclination angle $\theta_f$ such that $\theta_f\!<\!\pi$, 
when they are spun sufficiently rapidly.
This classification of  tippe tops into three groups and its classificatory criteria 
totally coincide with those obtained by Hugenholtz~\cite{Hugenholtz} and 
Leutwyler~\cite{Leutwyler}, both of whom resorted to completely different arguments and methods.

In the latter half of this paper we study the steady states  for spinning motion of the tippe top and 
examine their stability (or instability). 
It is well understood  that  the main source for the tippe top inversion 
is sliding friction~\cite{Braams,Hugenholtz},  which depends on the slip velocity $\vv_P$ 
of the contact point between the tippe top and  a table.
Often used is  Coulomb friction (see Eq.(\ref{Coulomb})). In fact, Coulomb friction 
is practical when $\vert \vv_P  \vert$ is away from zero,  but it
is  undefined for $\vv_P\!=\!0$.  However, we learn that at the steady state of the tippe top, 
the slip velocity $\vv_P$  necessarily vanishes.  
In order to facilitate a linear stability analysis of steady states and also to 
study the motion of the tippe top as realistically as possible, we adopt in our analysis a modified version of 
Coulomb friction   (see Eq.(\ref{Friction})), which is continuous in $\vv_P$ and vanishes 
at $\vv_P\!=\!\zerov$. 

Actually the steady states of the tippe top and their stability were
analyzed by Ebenfeld and Scheck~[ES]~\cite{EbenfeldScheck}, who assumed a similar  frictional force 
which is continuous at $\vv_P\!=\!\zerov$.  
They used the total energy of
the  spinning top as a Liapunov function. The steady states were found as solutions 
of constant energy.  And the stability or instability of these states were judged by examining
whether the Liapunov function assumes a minimum or  a maximum  at these states. 
Also recently, Bou-Rabee, Marsden and Romero [BMR]~\cite{BMR} analyzed
the tippe top inversion as a dissipation-induced  instability and,  using the modified Maxwell-Bloch equations 
and an energy-momentum argument,
they gave criteria  for the stability of the non-inverted and inverted states of the tippe top. 

We take a different approach to this problem. First, in order to  
find the steady states for spinning motion of the tippe top, we follow the method used by Moffatt, 
Shimomura and Branicki [MSB] for the case of  spinning spheroids~\cite{MSB}. Then   
the stability of these steady states is examined 
as follows:  Once a steady state is known, the system is perturbed around the steady state. 
Particularly we focus our attention  on the 
variable $\theta$, which is perturbed to $\theta=\theta_s+\delta\theta$, 
where $\theta_s$ is a value at the steady state and   $\delta\theta$ is a small quantity. 
Using the equations of motion, we obtain,  under the linear approximation,
a  first-order ODE for $\delta\theta$ of the  form, 
$\delta {\dot \theta}=H_s\delta\theta$,   
where $H_s$  is expressed by the values of  dynamical variables at the steady state. 
Thus the change of $\delta\theta$ is 
governed by the sign of  $H_s$. If $H_s$ is positive (negative),  $|\delta\theta|$ will 
increase (decrease) with time. Therefore, we conclude that 
when  $H_s$ is negative (positive), then the state is stable (unstable).
Using this new and rather intuitive criterion we argue about the stability of the steady states 
in terms of the initial spin velocity $n_0$ given at the position near $\theta \!=\!0$. 
We observe that our results on the stability of the steady states are consistent with ones 
obtained by ES and MSB. 
Then we obtain a critical value $n_c$ of the initial spin  which is required for the tippe top of Group II 
to flip over  up to the completely inverted position at $\theta=\pi$. 
Finally we confirm by simulation our  results on  the relation between the initial spin $n_0$ 
and the stability  of the steady states.

The paper is organized as follows: In Sec.~2 we explain the notation and  geometry used in this paper, and give 
all the necessary equations for the analysis of the spinning motion of the tippe top. 
In Sec.~3 we discuss about the GBC and its relevance to the inversion
phenomenon of the tippe top. We also show that the assumption of the GBC leads to  
the classification of tippe tops into three groups. 
Then in Sec.~4 we study the steady states for the spinning motion of the tippe top 
and examine their stability. Sec. 5 is devoted to a summary and discussion. 
In addition, we present four appendices. In Appendix A, the equations of motion which are used to analyze the
spinning motion of the tippe top are  enumerated.  In Appendix B, it is shown that 
 intermediate steady states for the tippe tops of Group II and Group III are stable when 
an initial spin $n(\theta\!=\! 0)$ falls in a certain range. In Appendix C we demonstrate 
that  our stability criterion for the steady state is equivalent to the one obtained by ES. 
And finally, in Appendix D, we show that our results on the stability of the vertical spin states 
are consistent with the  criteria derived by BMR.

\vspace{1cm}
\section{Equations of motion for tippe tops}
\smallskip

A commercially available tippe top is  usually a truncated sphere with a cylindrical stem. 
Instead  we perform our analysis taking a loaded (eccentric) sphere version of the tippe top. 
The center of mass is off center by a distance $a$. There are no qualitative differences 
between the two.  But if applied to the case of a commercial tippe top with a stem,  
our assertions would be valid up to the point when the stem touched the table surface.


\begin{figure}
  \begin{center}
    \includegraphics[width=0.6\textwidth]{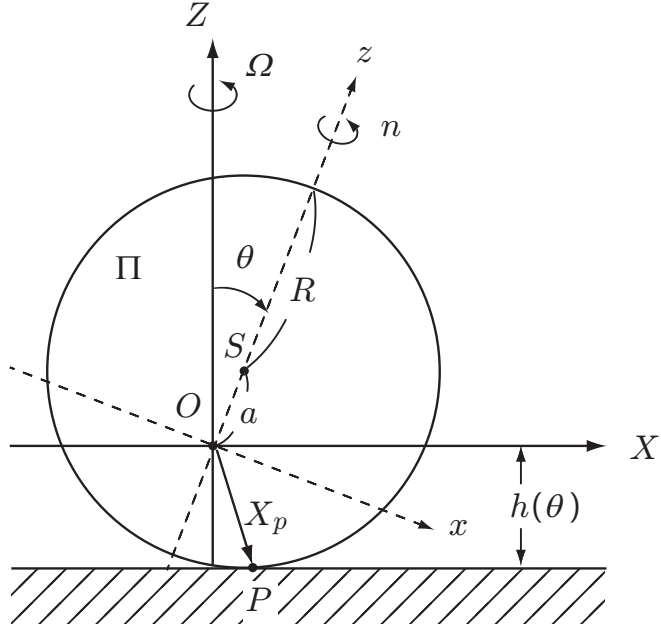}
    \caption{
      \label{LoadedSphere} A loaded sphere (eccentric) version of the tippe top.
      The center of mass $O$ is off center ($S$) by distance $a$.
      The tippe top spins on a horizontal table with point of contact $P$. Its axis of symmetry,
      $O\!z$, and the vertical axis, $O\!Z$, define a plane $\Pi$,  which precesses about $O\!Z$
      with angular velocity  \mbox{\boldmath$\Omega$}$(t)=(0,0,\Omega)$.
      $O\!X\!Y\!Z$ is a rotating frame of reference  with $O\!X$ horizontal in the plane $\Pi$.
      The height of $O$ above the table is $h(\theta)=R\!-\!a{\rm cos}\theta$, where $R$ is the radius.
      The position vector of $P$ from $O$ is $\Xv_P=(X_P,0,Z_P)$, where $X_P=\frac{dh}{d\theta}$ and $Z_P=-h(\theta)$.
    }
  \end{center}
\end{figure}

Fig.~\ref{LoadedSphere} shows the geometry.  
An axisymmetric tippe top spins on a horizontal table with point of contact $P$. 
We will work in a rotating frame of reference $O\!X\!Y\!Z$, 
where the center of mass is at the origin, $O$.  The center $S$ of the sphere with
radius $R$ is at a distance $a$ from the origin.
The symmetry axis of the tippe top, $O\!z$, and the vertical axis, $O\!Z$, define a plane 
$\Pi$,  which precesses about $O\!Z$ with angular velocity 
\mbox{\boldmath$\Omega$}$(t)=(0,0,\Omega)$. 
Let $(\phi, \theta, \psi)$ be the Euler angles of the body relative to $OZ$. Then we have 
$\Omega=\dot \phi$,  where the dot represents
differentiation with respect to time,   and  $\theta$  is the angle between $O\!Z$ and $O\!z$. 
We choose the horizontal axis $O\!X$ in the plane $\Pi$ and thus $O\!Y$ is vertical to
$\Pi$ and inward. 

In a rotating frame of reference $O\!x\!y\!z$, where $O\!x$ is in the plane $\Pi$ and 
perpendicular to the symmetry axis $O\!z$ and where $O\!y$ coincides with $O\!Y$, 
the tippe top spins about $O\!z$ with the rate ${\dot \psi}$. Since 
$\Omegav$ is expressed as $\Omegav=-\Omega \sin\theta {\hat {\xv}} + 
\Omega \cos\theta {\hat {\zv}}$  in the frame $O\!x\!y\!z$,  
the angular velocity of the tippe top, $\omegav$, 
is given by $\omegav=-\Omega \sin\theta {\hat {\xv}} + {\dot \theta} {\hat {\yv}} +
n {\hat {\zv}}$. Here ${\hat {\xv}}$, ${\hat {\yv}}$, and ${\hat {\zv}}$ are unit vectors along 
$O\!x$, $O\!y$, and $O\!z$, respectively,  
$n(t)$ is given by $n=\Omega \cos\theta  + {\dot \psi}$. The $O\!x$ and $O\!y$ are not body-fixed axes but 
are principal axes, so that the angular momentum, $\Lv$, is expressed by 
$\Lv=-A\Omega \sin\theta {\hat {\xv}} + A{\dot \theta} {\hat {\yv}} + Cn {\hat {\zv}}$, 
where $(A,A,C)$ are the principal moments of inertia at $O$. 
Using the perpendicular axis theorem and the parallel axis theorem, we see that  
$A/C\ge\frac{1}{2}$  for any axisymmetric density distribution. 

The coordinate system $O\!x\!y\!z$ is obtained from the frame  $O\!X\!Y\!Z$ by rotating 
the latter about the $O\!Y$ ($O\!y$) axis through the angle $\theta$. Hence, in the 
rotating frame $O\!X\!Y\!Z$, $\omegav$ and $\Lv$ have components
\begin{eqnarray}
{\mbox{\boldmath$\omega$}}&=&\Bigl( (n-\Omega{\rm cos}\theta)\sin\theta, 
{\dot \theta}~,\Omega \sin^2\theta +n \cos\theta \Bigr)~, 
\label{AngularVelocity}\\
\Lv&=&\Bigl( (Cn-A\Omega{\rm cos}\theta)\sin\theta, 
A{\dot \theta}~,A\Omega \sin^2\theta +Cn \cos\theta \Bigr)~, \label{AngularEqs}
\end{eqnarray}
respectively.  The evolution of $\Lv$ is governed by Euler's equation
\begin{equation}
\frac{\partial\Lv}{\partial t}+\Omegav\times 
\Lv =\mbox{\boldmath$X$}\!_P \times
(\mbox{\boldmath$N$}+\mbox{\boldmath$F$})~, \label{Euler}
\end{equation}
where $\mbox{\boldmath$X$}\!_P$ is the position vector of the contact point $P$ from $O$, 
$\mbox{\boldmath$N$}$ is the normal reaction at $P$, $\mbox{\boldmath$N$}=(0,0,N)$, 
with $N$ being of order $M\!g$, the weight, and  
$\mbox{\boldmath$F$}=(F_X, F_Y, 0)$ is the frictional force at $P$. 
We consider only {\it the situation in which the tippe top is always in contact with the table} 
throughout the motion.  
Since the point $P$ lies in the plane $\Pi$, $\mbox{\boldmath$X$}\!_P$ has 
components $(X\!_P, 0, Z\!_P)$, which are given by
\begin{subequations}
\begin{eqnarray}
Z_P&=& -(R-a\cos\theta)\equiv -h(\theta)~,\label{height.a}\\
X\!_P&=&a\sin\theta~=\frac{dh}{d\theta}, \label{height.b}
\end{eqnarray}
\end{subequations}
where $h(\theta)$ is the height of $O$ above the table.  
The components of (\ref{Euler}) are expressed, respectively, as
\begin{subequations}
\begin{eqnarray}
{\dot L}_X-\Omega L_Y&=&h(\theta)F_Y~,  \label{EulerX}\\
{\dot L}_Y+\Omega L_X&=&-a\sin\theta N- h(\theta)F_X ~,  \label{EulerY}\\
{\dot L}_Z&=&a\sin\theta F_Y~.  \label{EulerZ}
\end{eqnarray}
\end{subequations}
In terms of $\theta$, $\Omega$, and $n$ the above equations are rewritten  as
\begin{subequations}
\begin{eqnarray}
A{\dot \Omega}\sin\theta&=&(Cn-2A\Omega\cos\theta){\dot \theta}
+(a-R\cos\theta)F_Y~,  \label{EulerOmega}\\
A{\ddot \theta}&=&-\Omega(Cn-A\Omega\cos\theta)\sin\theta
-a\sin\theta N- h(\theta)F_X ~,  \label{Eulertheta}\\
C{\dot n}&=&R\sin\theta F_Y~.  \label{Eulern}
\end{eqnarray}
\end{subequations}

Now it is easily seen  from (\ref{AngularEqs}), (\ref{EulerX}) and (\ref{EulerZ}) 
that there exists an exact constant of  motion, 
\begin{equation}
J=-\mbox{\boldmath$L$}\cdot \mbox{\boldmath$X$}\!_P
=-L_X\frac{dh}{d\theta}+L_Z h(\theta)~\qquad ({\rm a\ constant}),
\label{JellettConstant}
\end{equation}
which is valid irrespective of the reaction force
$(\mbox{\boldmath$N$}+\mbox{\boldmath$F$})$  at the contact point $P$, in other words, 
whether or not slipping occurs.  
This so-called ``Jellett's  constant"~\cite{Jellett} is  typical for the tippe top 
whose portion of the surface in contact with the table is spherical.

The velocity, $\vv_{{\rm rot}P}$, of the contact point $P$ with respect to the center of mass
$O$ is given by 
$\vv_{{\rm rot}P}=\omegav  \times \mbox{\boldmath$X$}\!_P$,  and thus has components,  
\begin{subequations}
\begin{eqnarray}
v_{{\rm rot}PX}&=&-h(\theta)\dot \theta~, \label{vPX} \\
v_{{\rm rot}PY}&=&\left\{ R(n-\Omega  \cos\theta )+a \Omega\right\}
\sin\theta~, \label{vPY}  \\ 
v_{{\rm rot}PZ}&=&-a\sin\theta~\dot \theta~. \label{vPZ} 
\end{eqnarray}
\end{subequations}
The center of mass $O$ is not stationary.  
Let $\uv_O=(u_{OX}, u_{OY}, u_{OZ})$ represent the velocity of  $O$,
then  the slip velocity of the contact point $P$, $\vv_P=(v_{PX}, v_{PY}, v_{PZ})$,  is 
\begin{equation}
\vv_P=\uv_O+\vv_{{\rm rot}P}~. \label{VelP}
\end{equation}
Since $u_{OZ}=\frac{dh}{dt}=-v_{{\rm rot}PZ}$~, we have $v_{PZ}=0$ as was expected.

The equation of motion for the center of mass $O$ is given by
\begin{eqnarray}
M \left( \frac{\partial \uv_O}{\partial t}+\Omegav \times \uv_O\right)=\Nv+\Fv+\Wv~,
\label{EqCM}
\end{eqnarray}
where $M$ is the mass of the tippe top and $\Wv=(0,0,-Mg)$ is the force of gravity. In components, 
Eq.(\ref{EqCM}) reads
\begin{subequations}
\begin{eqnarray}
M\left( {\dot u}_{OX}-\Omega u_{OY}\right)&=&F_X ~, \label{EqCMX}\\
M\left( {\dot u}_{OY}+\Omega u_{OX}\right)&=&F_Y ~,\label{EqCMY}\\
M{\dot u}_{OZ}&=&N-Mg~. \label{EqCMZ}
\end{eqnarray}
\end{subequations}
Since ${\dot u}_{OZ}=\frac{d^2 h}{dt^2}$, Eq.(\ref{EqCMZ}) gives
\begin{equation}
N=M\left\{ g+a\left( \dot \theta^2 \cos\theta+{\ddot \theta}~  \sin\theta 
\right) 
\right\}~,
\end{equation}
which shows that the normal force $N$ is  of order $Mg$ when
$a\dot \theta^2, ~a|{\ddot \theta}| \ll g$. 

We need an information on the frictional force $\Fv$. It is well understood that the 
sliding friction is the main source for the tippe top inversion~\cite{Braams,Hugenholtz}. 
So we will ignore other possible frictions, such as, rolling friction~\cite{KaneLevinson} and
rotational friction which is due to  pure rotation about a vertical axis .

Concerning the sliding friction, 
often used is a Coulomb law, which states that
\begin{equation}
\Fv_{\rm C}=-\mu N\frac{\vv_P}{\vert \vv_P  \vert}~. \label{Coulomb}
\end{equation}
where $\mu$ is a coefficient of friction. 
Another possibility is a viscous friction law, which states that the friction is 
linearly related to $\vv_P$. 
Coulomb friction is practical when $\vert \vv_P  \vert$ is away from zero but it is 
undefined at $\vv_P=0$.  The slip velocity of the  contact point $P$ 
necessarily vanishes at the steady state of the tippe top.  
In order to study the motion of the tippe top as realistically as possible 
and also to facilitate a linear stability analysis of steady states, we modify the expression 
of   Coulomb friction (\ref{Coulomb}) as 
\begin{equation}
\Fv=-\mu N\frac{\vv_P}{\vert \vv_P (\Lambda) \vert}~, \quad 
{\rm with} \quad \vert \vv_P (\Lambda) \vert=\sqrt{v^2_{PX}+v^2_{PY}+\Lambda^2}
~, \label{Friction}
\end{equation}
so that $\Fv$ is continuous in $\vv_P$ and vanishes at $\vv_P=\zerov$. 
Here we choose $\Lambda$ as a sufficiently small number with 
dimensions of velocity.
Note that  $v_{PZ}=0$ and thus the $Z$-component of $\Fv$ is 0.

This completes the presentation of all the necessary equations for the 
analysis of the motion  of tippe tops.  We enumerate all these equations  in Appendix A. 
We need further the initial conditions.
When we play with a tippe top, we usually  give it a rapid spin with its
axis of symmetry nearly vertical. So  let us choose the 
following initial conditions for $\theta$ and other angular velocities:
\begin{eqnarray}
\theta_0&=&\theta(t\!=\!0) \quad {\rm   small}~, \qquad 
{\dot \theta}_0={\dot \theta}(t\!=\!0)=0  \nn \\
\Omega_0&=&\Omega(t\!=\!0)=0,~\label{InitialRotation} \\  
\dot \psi_0 &=&\dot\psi(t\!=\!0)~ \quad {\rm large}~.\nn
\end{eqnarray}
We take~ $\theta_0\!=\!0.01\!\sim\! 0.1$ rad and $\dot \psi_0 \!=\!10\!\sim\!  150~ {\rm
rad/sec}$.  Recall that the spin $n(t)$  is given by 
$n=\Omega \cos\theta  + {\dot \psi}$, and thus we have $n_0\!=\!n(t\!=\!0)\!=\!10\!\sim\!  
150~ {\rm rad/sec}$.  
As for the initial condition for the velocity of the center of mass $O$, we take 
\begin{equation}
\uv_0=\uv_O(t\!=\!0)=\zerov~,  \label{InitialTrans}
\end{equation}
since we usually do not give a large translational 
motion to the tippe top at the beginning.

With the above initial conditions (\ref{InitialRotation}) and (\ref{InitialTrans}), 
we analyze  the behaviors of  the tippe top using 
three angular (\ref{EulerOmega}-\ref{Eulern}) and 
three translational (\ref{EqCMX}-\ref{EqCMZ}) equations of motion, together with the 
knowledge of the frictional force, a modified version of the Coulomb law  (\ref{Friction}), and 
the velocities (\ref{vPX}-\ref{vPZ}) and (\ref{VelP}). 
When we perform simulations we use the adaptive Runge-Kutta method.


\vspace{1cm}

\section{Gyroscopic balance condition}
\smallskip

\subsection {The variable $\xi$}
We define a variable $\xi$ as
\begin{equation}
\xi\equiv Cn-A\Omega \cos\theta~. \label{xi}
\end{equation}
In terms of $\xi$, the $X$- and $Z$- components of 
$\Lv$ in (\ref{AngularEqs}) 
and  Jellett's constant $J$, (\ref{JellettConstant}), are expressed, respectively,  as 
\begin{eqnarray}
L\!_X&=&\xi~\sin\theta,~ \qquad  L\!_Z=
\xi~ \cos\theta +A~\Omega~, \label{AngularEqs2} \\
J&=&-\xi a ~\sin^2\theta +L_Z h(\theta)~. \label{JellettConstant2}
\end{eqnarray}

The condition $\xi\!=\!0$ has been introduced by MS~\cite{MS} 
in their analysis of spinning hard-boiled eggs, and referred to as the GBC. They discovered that the GBC, $\xi\!=\!0$,  is
approximately  satisfied for the spinning egg and, using this GBC,  they resolved a long standing riddle: when a hard-boiled egg 
is spun  sufficiently rapidly on a table with its axis of symmetry horizontal, 
the axis will rise  from the horizontal to the vertical. 
We outline how MS found the GBC for the spinning egg~\cite{MS}. The system of  
the spinning egg obeys essentially the same equations of motion as the case of the 
tippe top, to be specific, Eqs. (\ref{Euler}) and (\ref{EqCM}). The $Y$-component of 
(\ref{Euler}) for the spinning egg is given by (\ref{Eulertheta}), with the factor,  
$a\sin \theta$,  being replaced by $X_P$. 
Because the secular change of $\theta$ is slow and
thus $|\ddot \theta|\ll \Omega^2$, the  term $A\ddot \theta$  
can be neglected.  Furthermore, in a situation where $\Omega^2$ is
sufficiently large so that the terms involving $\Omega$ in
(\ref{Eulertheta}) dominate the terms $-X\!_P N$ 
and $-h(\theta) F_X$, Eq.~(\ref{Eulertheta}) 
is reduced, in leading order,  to $(Cn-A\Omega \cos\theta)\Omega
\sin\theta=0$. Hence, for $\sin \theta\not= 0$, we arrive at the condition 
$\xi=Cn-A\Omega \cos\theta=0$.

The tippe top shows the similar behavior as the spinning egg.  Then one may ask:
does the GBC also hold for  the tippe top?   We will show that 
the answer is  ``partly no" and ``partly yes". ``Partly no" means that the GBC is 
not satisfied initially.  
Tippe tops are usually spun  with 
$\theta_0\approx 0$, $\Omega_0\approx 0$, and  large $\dot \psi_0$ and, therefore,
$n_0 \approx\dot \psi_0$ is large, from which we find that  $\xi_0\!=\!\xi(t\!=\!0)
\approx Cn_0$ is large\footnote{In this paper we always take the initial spin velocity $\dot \psi_0$ 
about $Oz$ to be  positive and, therefore, 

\quad $\xi_0$ is positive.}.  
Thus the GBC does not hold at the beginning. 
%
%
However, we will see later that the GBC does approximately hold whenever the tippe top rises, 
which is the meaning of ``partly yes". 
In fact,  the argument of MS to derive the GBC for the spinning egg  
can also  be applied to the tippe top.  Thus in a situation where $\Omega$ is sufficiently large 
and for $\sin \theta\not= 0$, the GBC is expected to be satisfied.  
On the other hand, in the case of  the  spinning egg, 
the GBC is approximately satisfied initially. 
We start to spin an egg with its symmetry axis horizontal,
that is, with $\theta_0\approx \frac{\pi}{2}$, $\dot \psi_0\approx 0$ and large $\Omega_0$. 
Hence we find $n_0\approx 0$ and $\xi_0\approx 0$ for the spinning egg. 

We emphasize that  the variable $\xi$ initially takes  a large positive value 
for the tippe top. But our numerical
analysis   will show  that  when a tippe top turns over,  $\xi$ soon makes  a  rapid transition from large positive values to 
negative values and 
%
%
starts to oscillate about a small negative value.

Before proceeding with a discussion of this transition of  $\xi$,  let us consider 
the consequences when  the GBC is exactly satisfied  for the tippe
top.

\subsection{Consequences of the exact GBC}

In a situation where $\Omega$ is sufficiently large and $\theta$ is not in the vicinity 
of 0 or $\pi$, the GBC is  realized for the tippe top.
Let us consider the case  that the exact GBC, $\xi=0$, is satisfied for the tippe top.  
Then,  we have
\begin{equation}
J=L_Z h(\theta)~,  \label{GBCJellett}
\end{equation}
from (\ref{JellettConstant2}),  and $L_Z=A\Omega$ from the second equation in 
(\ref{AngularEqs2}).  If the angular velocity $\Omega$ around the vertical axis 
is reduced and, therefore,  $L_Z$ decreases,  Eq.(\ref{GBCJellett}) 
tells us that the height $h(\theta)$ of the center of mass from the table 
increases since $J$ is a constant, which means the turning over of the  tippe top. 
Differentiating both sides of (\ref{GBCJellett}) by time 
and using (\ref{height.b}) and (\ref{EulerZ}), 
we obtain  
a first-order ODE for $\theta$, 
\begin{equation}
J\dot\theta=-F_Y h^2(\theta)~. \label{EquForTheta}
\end{equation}

We  assume also that the $Y$-component of $\uv_O$, 
the translational velocity of the center of mass $O$, 
in (\ref{VelP}) is negligible
in the first approximation as compared with that of $\vv_{{\rm rot}P}$,  and 
we set $v_{PY}=v_{{\rm rot}PY}$. We  see that 
numerical simulation supports this assumption. 
Then, one can use Eq. (\ref{vPY}) and the GBC to eliminate $n$ and $\Omega$, and obtain 
$v_{PY}$ as only a function of the dynamical variable $\theta$ as follows: 
\begin{equation}
v_{PY}=\frac{J\sin\theta}{Ah(\theta)}\left\{ a+R\Bigl(\frac{A}{C}-1 \Bigr)
\cos\theta \right\}~. \label{Vpsecond}
\end{equation}
Since the frictional force $F_Y$ is proportional to $v_{PY}$, 
we obtain from  Eqs.(\ref{EquForTheta}-\ref{Vpsecond}),  
\begin{equation}
\dot\theta \propto \widetilde v_{PY}  \label{EquTheta}
\end{equation}
with a {\it positive} proportional coefficient and 
\begin{equation}
\widetilde v_{PY}=\sin\theta \left\{ a+R\Bigl(\frac{A}{C}-1 \Bigr)
\cos\theta \right\}~.  \label{VPYTilde}
\end{equation}
Equation (\ref{EquTheta}) implies that 
the change of $\theta$  is governed by 
the sign of $\widetilde v_{PY}$. If $\widetilde v_{PY}$ is positive (negative), 
then $\theta$  will increase (decrease) with time.
Therefore a close examination of the behavior of $\widetilde v_{PY}$ as a function of 
$\theta$  will be important \footnote{A resemblance of (\ref{EquTheta}) to a renormalization group equation
which appears in quantum field theories for critical phenomena and high energy physics
is emphasized in Sec. 5.
}. 

We observe from (\ref{VPYTilde}) that $\widetilde v_{PY}\!=\!0$ at $\theta\!=\!0$ and $\pi$,  
since $\sin \theta \!=\!0$ at these angles. Moreover,  
$\widetilde v_{PY}$ may vanish at  an other angle,  which is given by solving
\begin{equation}
a+R\Bigl(\frac{A}{C}-1 \Bigr) \cos\theta =0~. \label{VPzero}
\end{equation}
Equation (\ref{VPzero}) has a solution for $\theta$ if $\frac{A}{C}<1\!-\!\frac{a}{R}$ or 
$1\!+\!\frac{a}{R}<\frac{A}{C}$ and  no solution otherwise. Accordingly,  
tippe tops are classified into three groups, depending on the values of $\frac{A}{C}$ and 
$\frac{a}{R}$:  Group I with $\frac{A}{C}<1\!-\!\frac{a}{R}$;  
Group II with $1\!-\!\frac{a}{R}<\frac{A}{C}<1\!+\!\frac{a}{R}$;  and Group III with
$1\!+\!\frac{a}{R}<\frac{A}{C}$.  

\bigskip

We now examine the behaviors of tippe tops belonging to each group.

\begin{figure}
  \begin{center}
    \begin{tabular}{cc}
      \includegraphics[width=0.46\hsize]{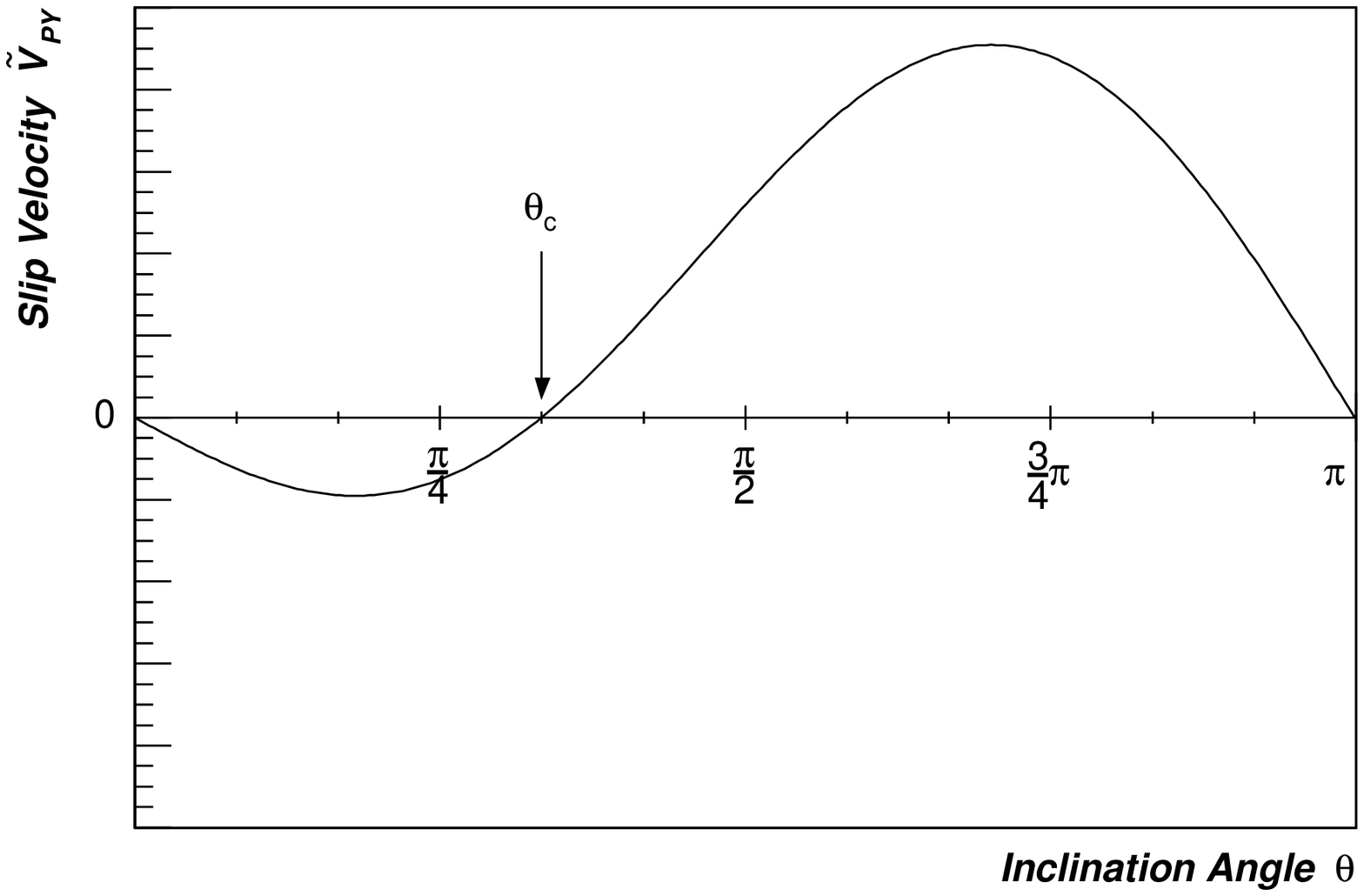} &
      \includegraphics[width=0.46\hsize]{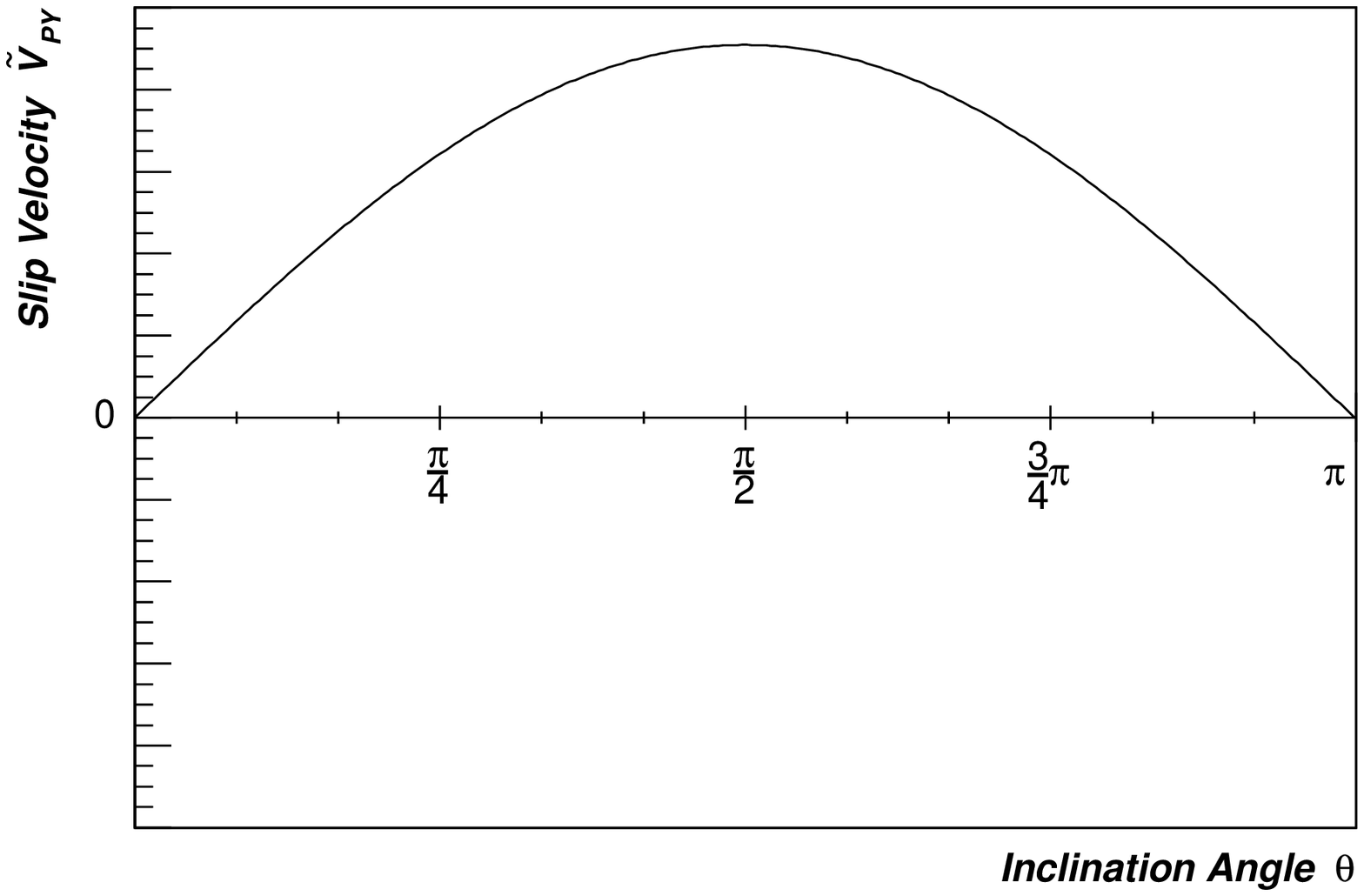} \vspace{-10pt} \\
      (a) & (b) \\
      \includegraphics[width=0.46\hsize]{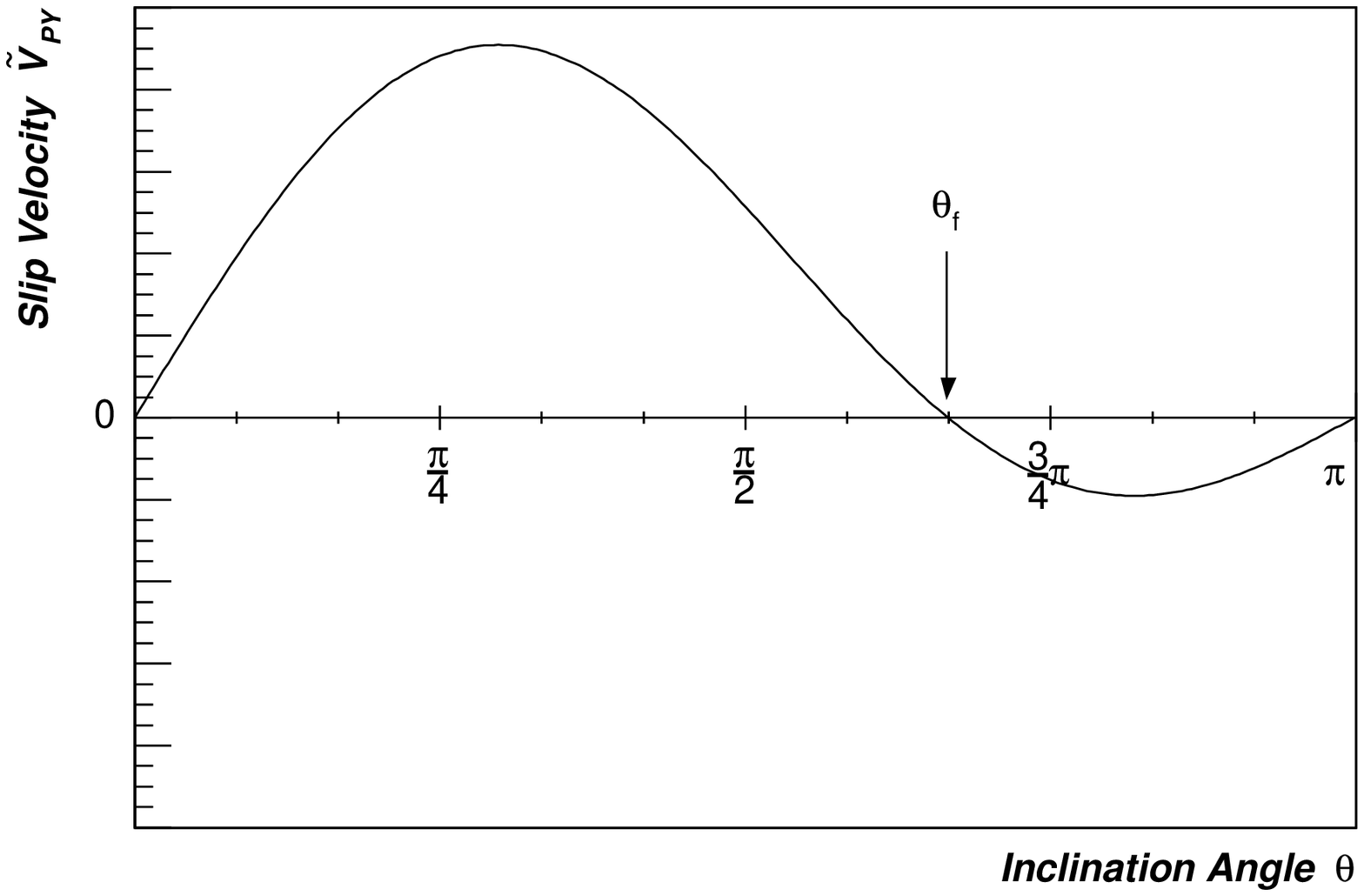} & \vspace{-10pt} \\
      (c) & \\
    \end{tabular}
    \caption{
      $\widetilde V_{PX}$ as a function of $\theta$ for tippe tops of
      (a) Group I with $\frac{a}{R}=0.1$ and $\frac{A}{C}=0.8$; 
      (b) Group II with $\frac{a}{R}=0.1$ and $\frac{A}{C}=1$; 
      (c) Group III with $\frac{a}{R}=0.1$ and $\frac{A}{C}=1.2$.
    }
    \label{VpGroups}
  \end{center}
\end{figure}

\noindent
(i) Group I \quad ($\frac{A}{C}<1\!-\!\frac{a}{R}$ )\\
Imagine that a familiar top consisting of a circular disk and a stem is 
located inside of a hollow massless sphere. The stem is along the diameter 
of the sphere whose center does not coincide with the center of mass.
This toy may belong to  Group~I. 
Figure~\ref{VpGroups} (a) shows a typical graph of $\widetilde v_{PY}$ for  a tippe top of 
Group I.   The graph crosses the line $\widetilde v_P=0$ at an angle 
\begin{equation}
\theta_c=  \cos^{-1} \left( \frac{a}{R(1-\frac{A}{C})}  \right)
\quad {\rm and} \quad 0<\theta_c<\frac{\pi}{2}~, \label{ThetaCritical}
\end{equation}
and 
$\widetilde v_{PY}$ is positive for $\theta_c< \theta <\pi$ but negative for $0< \theta
<\theta_c$.  So, the angle $\theta_c$ is a {\it critical point}. If a tippe top of Group I 
is spun on a table  with $\dot \psi_0\!\approx\!0$ and sufficiently large $\Omega_0$ and with the
initial  angle $\theta_0>\theta_c$, then $\theta$ will increase to $\pi$, which means
that  the body   will eventually spin  at  $\theta\!=\!\pi$. In the case $\theta_0<\theta_c$, 
we will see  that the body spins  at  $\theta\!=\!0$.  Depending on the initial value 
$\theta_0$  the body will spin at  the end point $\theta\!=\!0$ or $\pi$.  
Both ends are stable points.  Usually we give a spin to the tippe top at a position with 
$\theta_0\approx0$. Spun at $\theta_0\approx0$, the tippe top of Group I does not turn over 
however large a spin it is given and 
 will stay
spinning at $\theta \approx 0$.

\bigskip

\noindent
(ii) Group II \quad ($1\!-\!\frac{a}{R}<\frac{A}{C}<1\!+\!\frac{a}{R}$ ) \\
Commercial tippe tops belong to Group II. A typical graph of $\widetilde v_{PY}$ for a tippe 
top of Group II is shown in Fig.~\ref{VpGroups} (b). 
We see that $\widetilde v_{PY}$ is positive for $0\!<\!\theta\!<\!\pi$.
Therefore, the end point at $\theta=0$ is unstable while the other end at $\theta=\pi$ is a stable
point.  Once given a sufficiently large spin at $\theta_0\approx0$, 
the tippe top of Group II will turn over and  spin at $\theta=\pi$. 
Actually, commercial tippe tops have stems. Thus for those tops the  above statement is 
valid up to the  angle when the stem touches the table.

\bigskip

\noindent
(iii) Group III \quad ($1\!+\!\frac{a}{R}<\frac{A}{C}$ )\\
For an example of the tippe top of Group III, we may imagine a 
prolate spheroid put inside of a hollow massless sphere. The symmetric axis is along the 
diameter of the sphere and the mass distribution is nonuniform so that 
the center of mass is apart from the sphere's center.
Figure~\ref{VpGroups} (c) shows a typical graph of $\widetilde v_{PY}$ for  a tippe top of Group III.  
Similarly to the  case of Group I the graph crosses the line $\widetilde v_P=0$ at an angle 
\begin{equation}
\theta_f=  \cos^{-1} \left( \frac{a}{R(1-\frac{A}{C})}  \right)
\quad {\rm and} \quad \frac{\pi}{2}<\theta_f<\pi~. \label{ThetaFixed}
\end{equation}
In this case 
$\widetilde v_{PY}$ is positive for $0< \theta <\theta_f$ and negative for $\theta_f< \theta
<\pi$.  So, both ends at $\theta=0$ and $\pi$ are unstable points,  while
the angle $\theta_f$ is a {\it fixed point}.   
When the body is spun sufficiently rapidly with the initial 
angle $\theta_0$ anywhere, $\theta$ will approach the fixed point $\theta_f$. 
Thus the tippe top of Group III, even though given a sufficiently large spin at $\theta_0\approx0$, 
will  never turn over to $\theta=\pi$ but up  to the angle $\theta_f$. 

Now it should be emphasized that the argument so far for the classification of tippe tops into 
three groups is based on the assumption that the  GBC, $\xi=0$, is exactly satisfied.  
It is very interesting to note 
that the above classification into three groups and its classificatory criteria 
totally coincide with those obtained by Hugenholtz~\cite{Hugenholtz} and 
Leutwyler~\cite{Leutwyler}, both of whom resorted to completely different 
arguments and methods. In fact, Hugenholtz  considered the effect on the tippe top when 
a small frictional force is working during the uniform motion and  reached the same conclusion. 
On the other hand,  Leutwyler used Lagrangian formalism and  searched for the minimum of  
energy for the tippe top under the constraint of  Jellett's constant
 (\ref{JellettConstant}). Finally the behavior of the tippe top under the 
GBC was studied earlier by Sakai \cite{Sakai}. Unfortunately, his work was written 
in Japanese and is, therefore, not well known. The consequences derived in this subsection 
partly overlap with his results.

\subsection{The behavior of  the variable $\xi$}
\begin{figure}
  \begin{center}
    \begin{tabular}{cc}
      \includegraphics[width=0.49\textwidth]{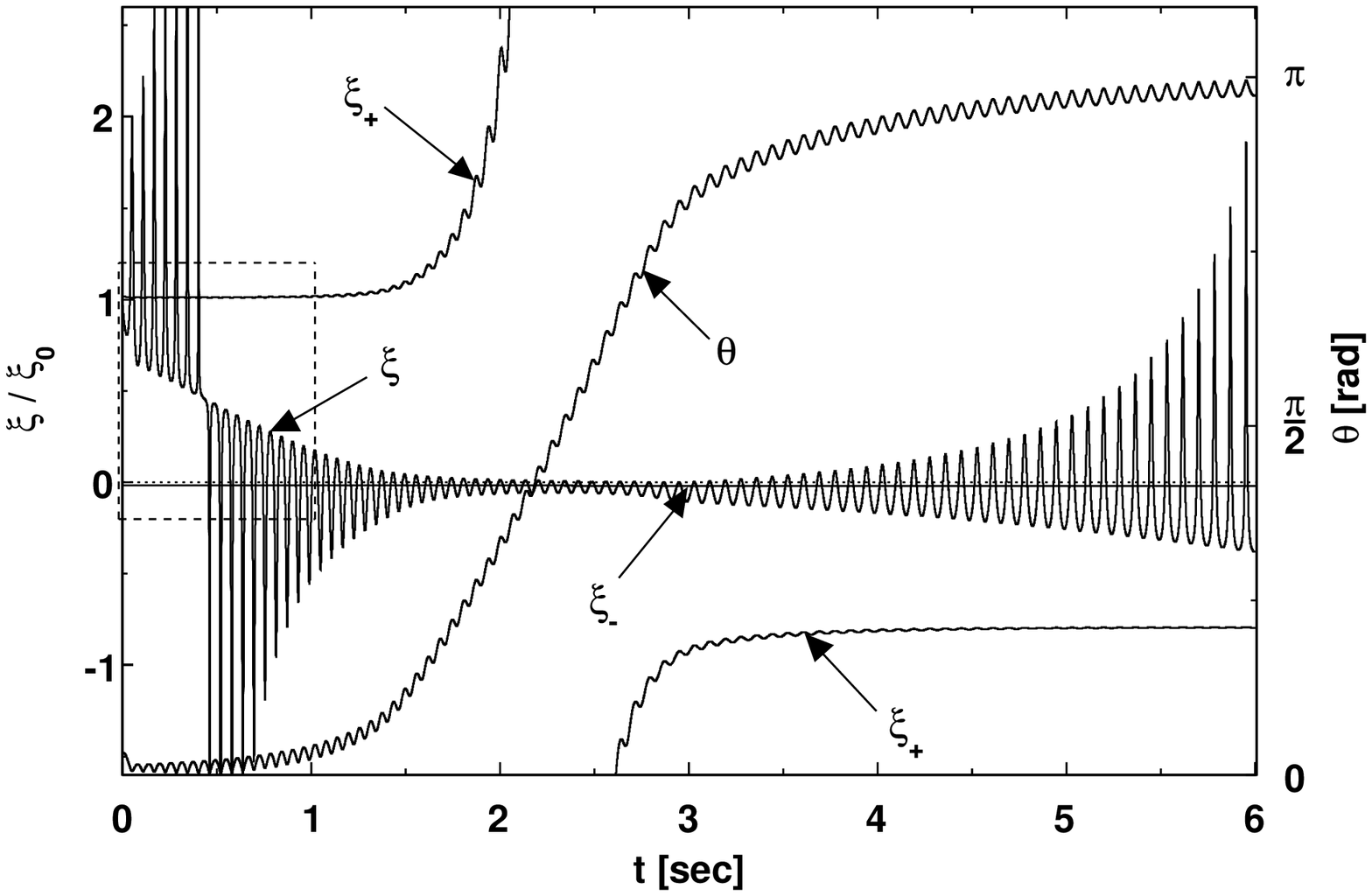} &
      \includegraphics[width=0.49\textwidth]{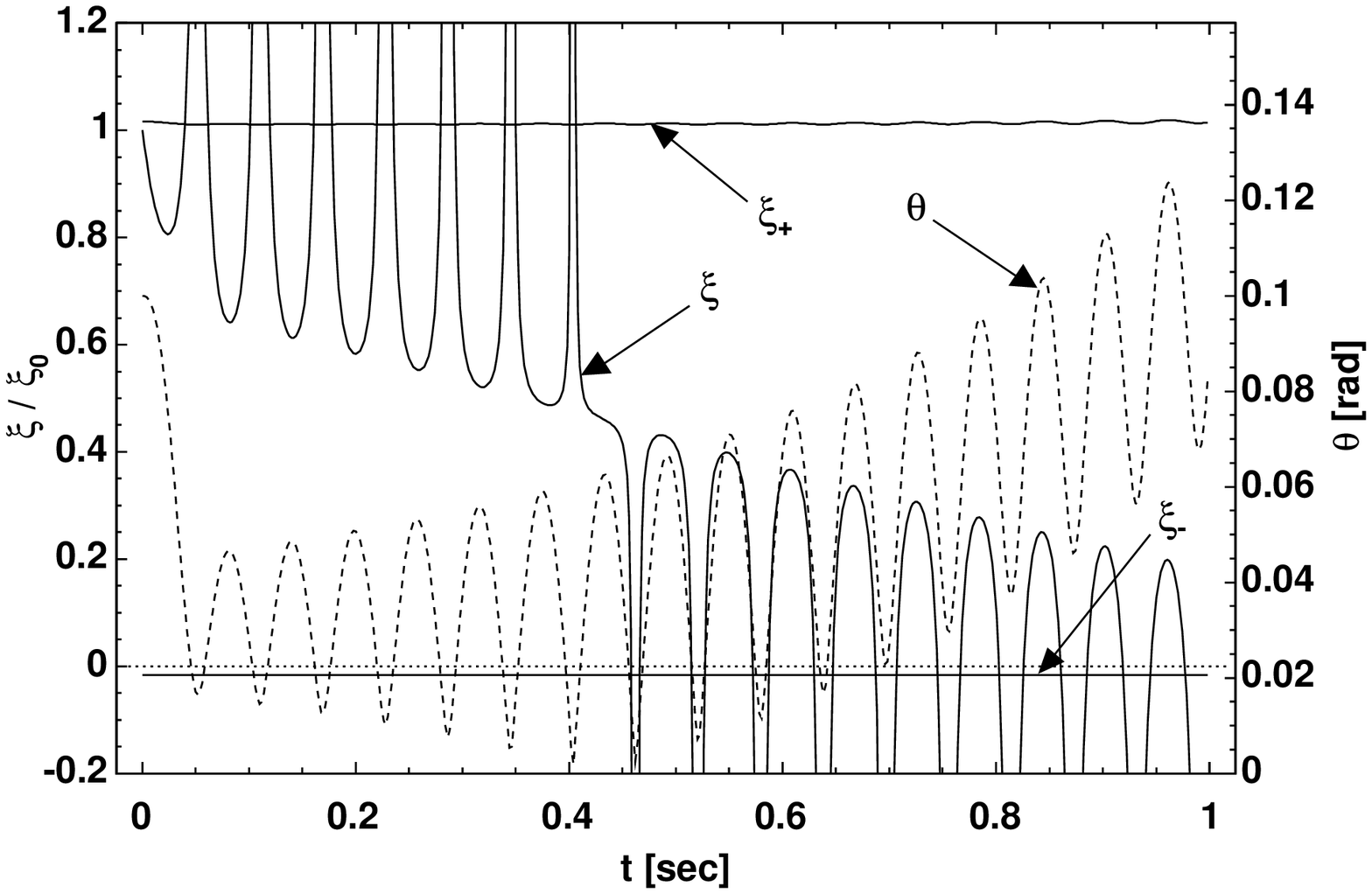} \\
      (i) & (ii)
    \end{tabular}
    \caption{
      \label{BehaviorXi(a)}
      (i) Time evolution of the variable $\xi$ and the inclination angle $\theta$. Input parameters are 
      $R\!=\!1.5$ cm, $a\!=\!0.15$ cm, $M\!=\!15$ g, $g\!=\!980$ cm/sec${}^2$, $A\!=\!C\!=\!2/5 M R^2$,
      $\mu\!=\!0.1$,
      $\Lambda\!=\!1$ cm/sec.   Initial conditions are $\theta_0\!=\!0.1$ rad, $n_0\!=\!100$ rad/sec,
      $\dot{\theta}_0\!=\!\Omega_0\!=\!0$,
      $\uv_0\!=\!\zerov$. The curves $\xi_{\pm}$ are given by 
      Eq.(\ref{xiSolution}).
      (ii)  Blow-up of the section surrounded by dashed lines in (i). 
    }
    \vspace{1cm}
    \begin{tabular}{cc}
      \includegraphics[width=0.49\textwidth]{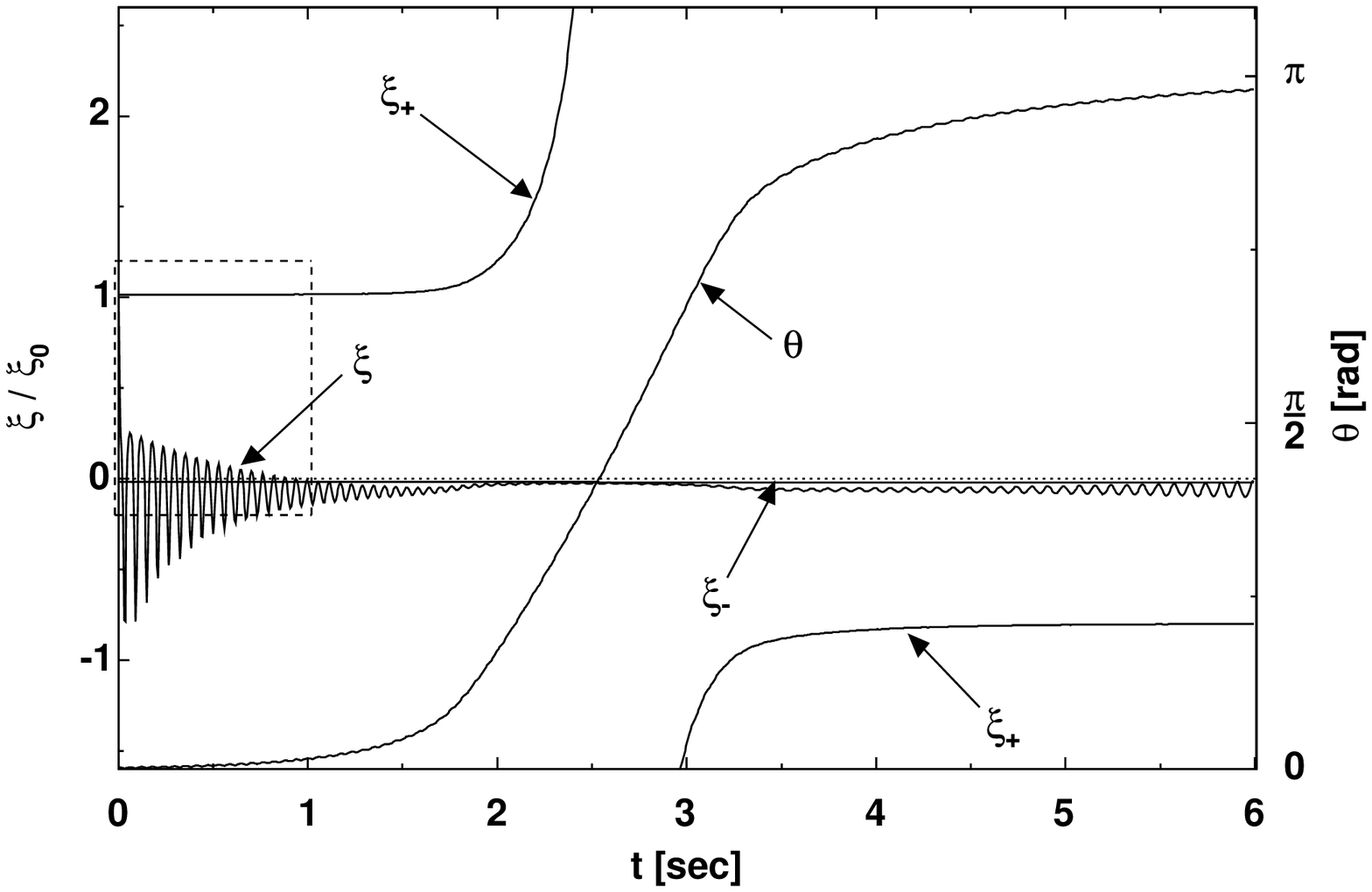} &
      \includegraphics[width=0.49\textwidth]{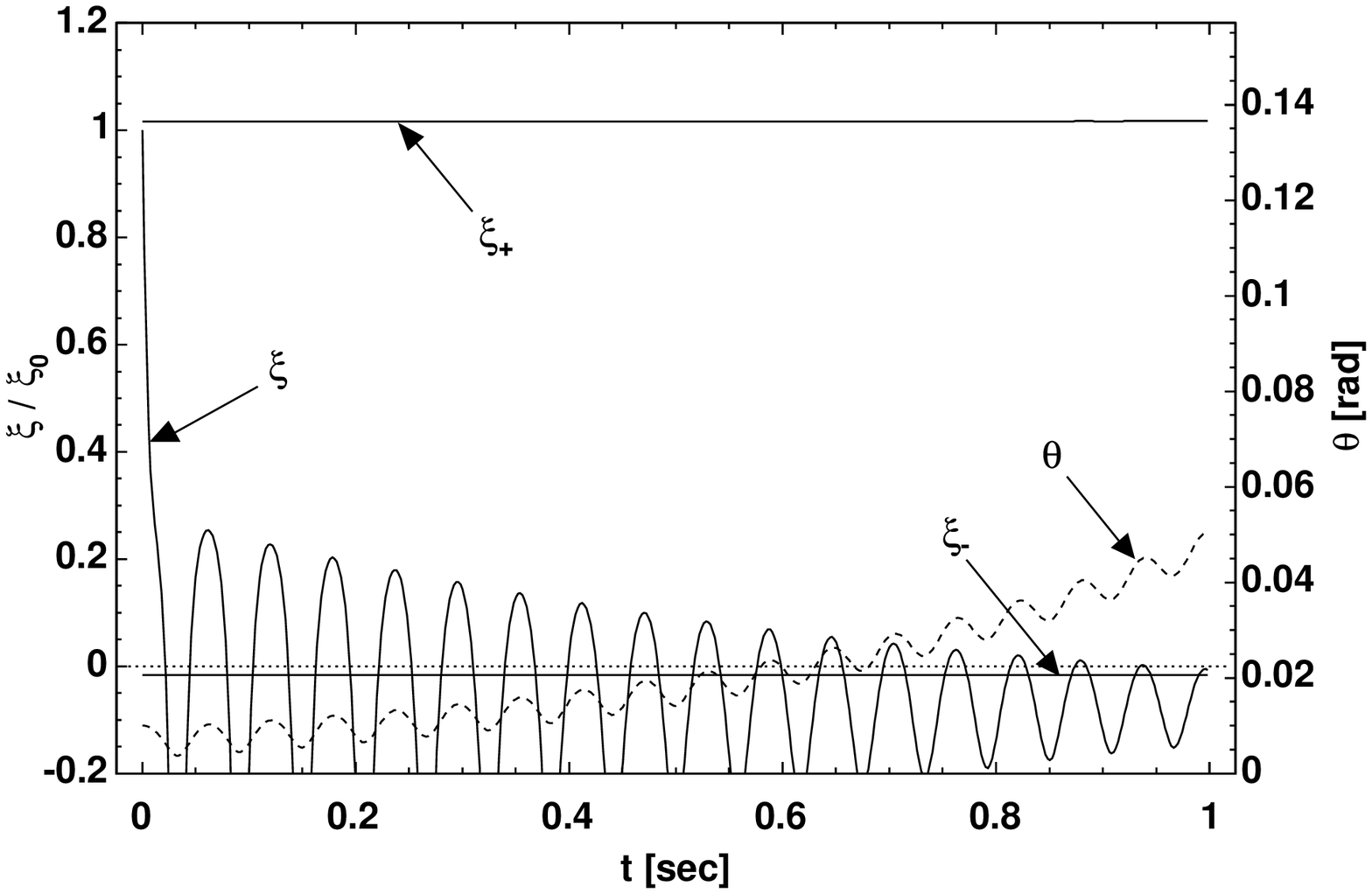} \\
      (i) & (ii)
    \end{tabular}
    \caption{
      \label{BehaviorXi(b)}
      (i) Time evolution of the variable $\xi$ and  the inclination angle $\theta$ with an initial 
      condition $\theta_0\!=\!0.01$ rad. Input parameters and other initial conditions are the same as 
      in Figure \ref{BehaviorXi(a)}.
      The curves $\xi_{\pm}$ are given by 
      Eq.(\ref{xiSolution}).
      (ii)  Blow-up of the section surrounded by dashed lines in (i). 
    }
  \end{center}
\end{figure}

\begin{figure}
  \begin{center}
    \begin{tabular}{cc}
      \begin{minipage}{0.4\hsize}
        \begin{center}
          \includegraphics[width=0.99\hsize]{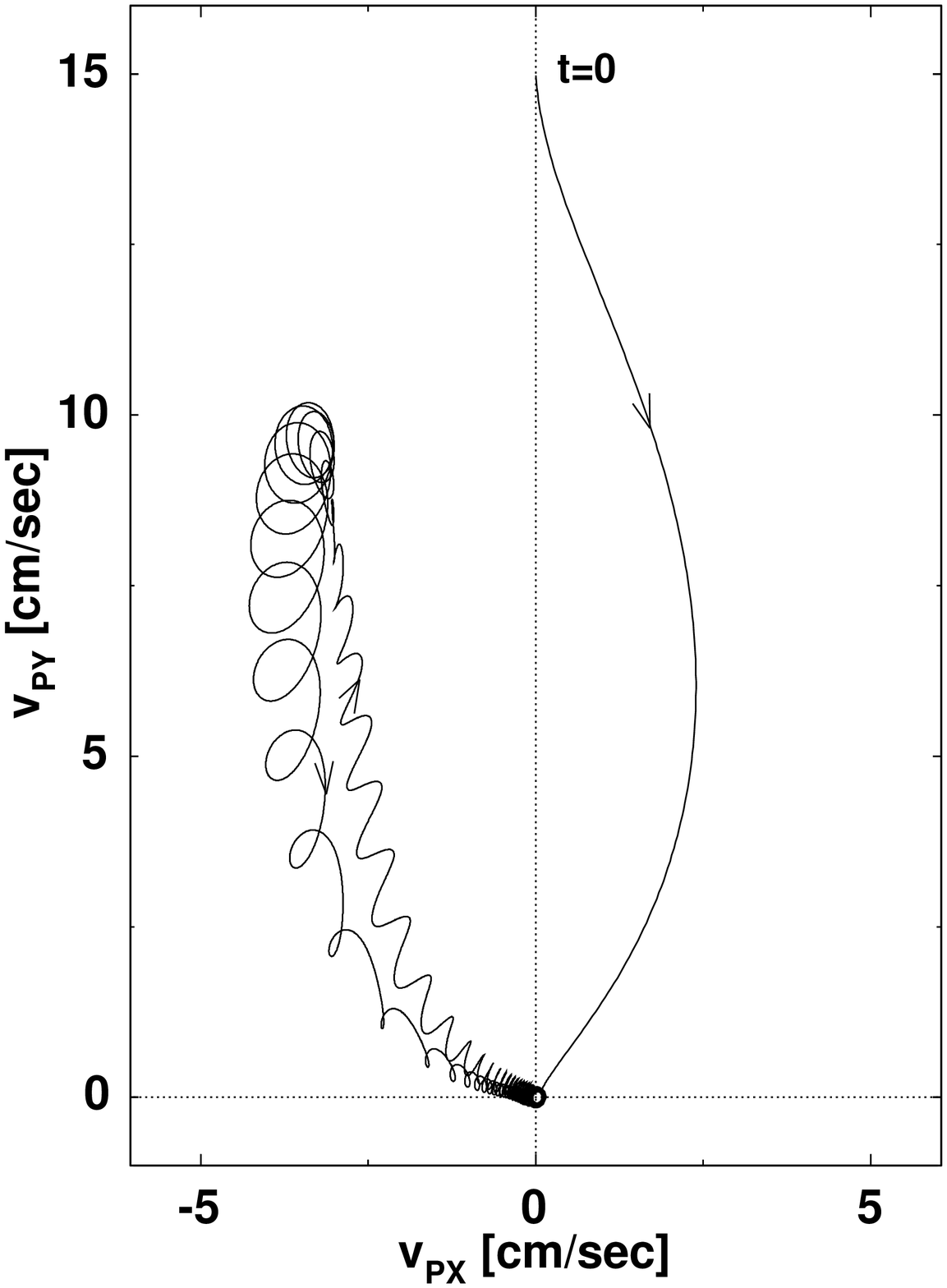}
        \end{center}
      \end{minipage} &
      \begin{minipage}{0.4\hsize}
        \begin{center}
          \includegraphics[width=0.99\hsize]{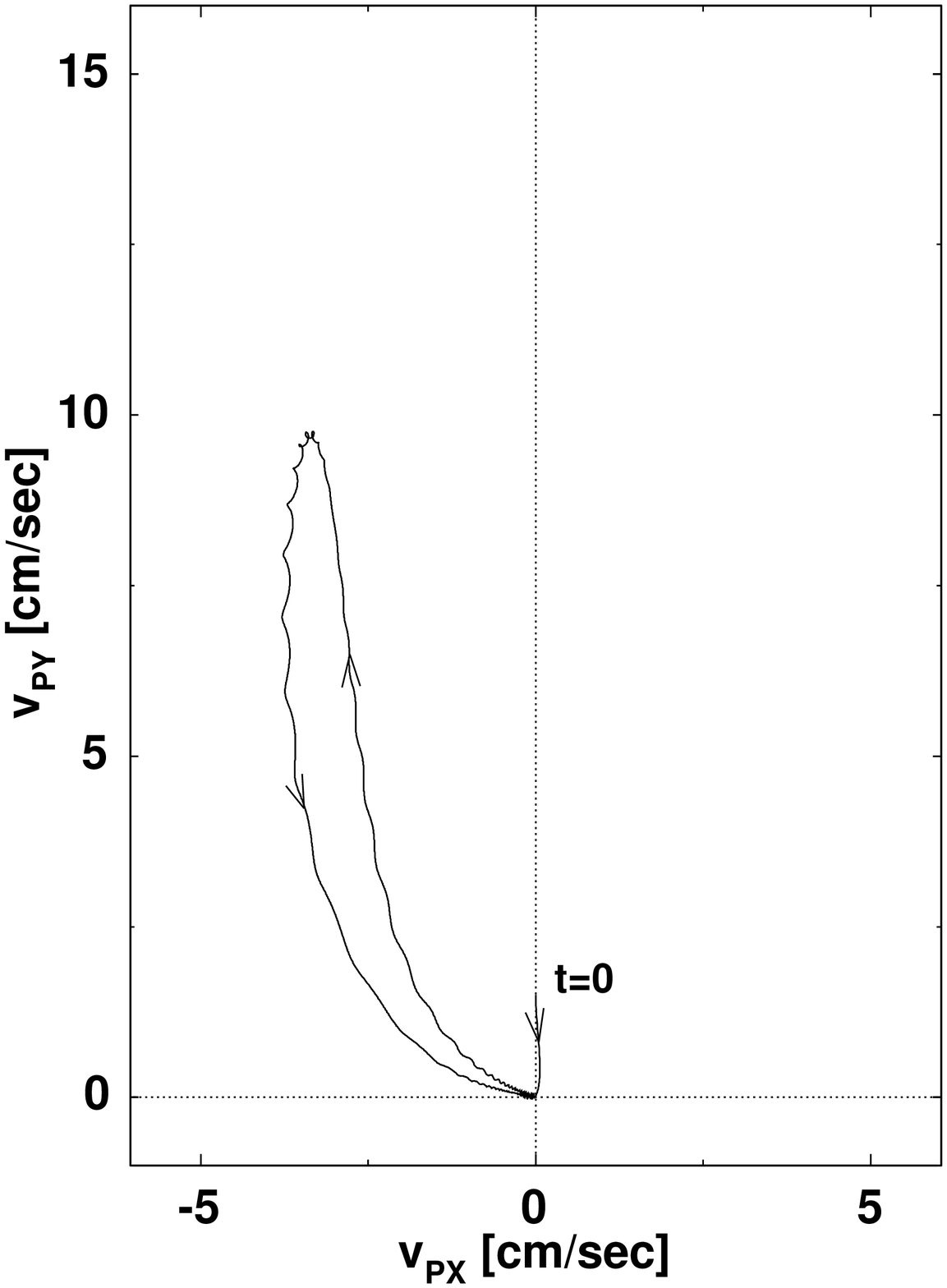}
        \end{center}
      \end{minipage} \\
      (a) & (b)
    \end{tabular}
    \caption{
      \label{TraVelocity}
      The trajectories of  slip
      velocity  of the contact point $P$ in the $(v_{PX}, v_{PY})$ space which are 
      obtained from (a)  the simulation shown in Fig.\ref{BehaviorXi(a)} with $\theta_0= 0.1$ and (b) 
      the one in Fig.\ref{BehaviorXi(b)} with $\theta_0= 0.01$.
    }
  \end{center}
\end{figure}

 As stated before, the GBC is not satisfied  initially for the tippe top. Actually, the initial value of  
$\xi$ is large and  positive. We have performed numerical computations 
to see the behaviors of $\xi$ and $\theta$ in time $t$. 
Typical examples are shown in Figs. \ref{BehaviorXi(a)} and \ref{BehaviorXi(b)}, where  
the scale of the left  sides is for $\xi$ normalized by the initial value $\xi_0$, while the scale of 
the right sides is for $\theta$ in radian.  Input parameters are for  both cases 
\begin{eqnarray}
R&=&1.5~{\rm cm}, \quad a=0.15~{\rm cm}, \quad M=15~{\rm g}, \quad g=980 ~{\rm cm/sec}^2, \nn
\\ A&=&C=\frac{2}{5} M R^2, \quad  \mu=0.1~, \quad \Lambda=1{\rm cm/sec}.\label{InputParameter}
\end{eqnarray}
For initial conditions we choose $n_0\!=\!100~ {\rm rad/sec}, ~\dot{\theta}_0\!=\!\Omega_0\!=\!0$, and  
$\uv_0\!=\!\zerov$ for both cases, but  we take $\theta_0\!=\! 0.1$ rad 
for the simulation shown in Fig.\ref{BehaviorXi(a)}  
and $\theta_0\!=\! 0.01$ rad for the one in Fig.\ref{BehaviorXi(b)}. 
With these initial conditions we have $\xi_0=Cn_0$ and $J=\xi_0(R\cos\theta_0-a)$. 
We show in Fig.\ref{TraVelocity} the trajectories of  slip
velocity  of the contact point $P$ in the $(v_{PX}, v_{PY})$ space which are 
obtained from the above simulations with (a) $\theta_0\!=\! 0.1$ and (b) $\theta_0\!=\! 0.01$.
The argument in section 3.2 tells us that 
a tippe top represented by the  input parameters (\ref{InputParameter})
is classified into  Group II and, therefore,   
this tippe top would turn over up to an inverted position, $\theta\!=\!\pi$, when it is given a 
sufficiently large initial spin. 


Also plotted in Figs.\ref{BehaviorXi(a)} and \ref{BehaviorXi(b)} are the curves $\xi_+$ and $\xi_-$, 
the expressions of which are given below. 
Using Eqs.(\ref{AngularEqs2}-\ref{JellettConstant2}) and replacing 
$\Omega$ and $L_Z$ with $\xi$, $\theta$ and $J$,  we find that 
the $X$- and $Y$- components of
the rotational equations (\ref{EulerX}-\ref{EulerY}) are rewritten as
\begin{subequations}
\begin{eqnarray}
{\dot \xi} \sin\theta&=&U(\xi, \theta, J){\dot \theta}+h(\theta)F_Y~, \label{EulerXX} \\
A{\ddot \theta}&=&-\frac{1}{Ah(\theta)}V(\xi, \theta, J)\sin\theta-h(\theta)F_X ~, \label{EulerYY} \\
{\rm with}      \hspace{2cm} && \nn\\ 
U(\xi, \theta, J)&=& \frac{J-\xi(R \cos \theta-a)}{h(\theta)}-\xi \cos\theta~, 
\label{EulerXXX} \\
V(\xi, \theta, J)&=&\left\{J-\xi(R \cos \theta-a)\right\}\xi 
+AMg a  h(\theta)~,
\label{EulerYYY} 
\end{eqnarray}
\end{subequations}
where we have set $N\!=\!Mg$. We see in Figs. {\ref{BehaviorXi(a)}} and {\ref{BehaviorXi(b)}},  
especially in the former, that $\theta$ changes while  nutating. 
We also see that the inflection points of  $\theta$, where the condition
$\ddot \theta=0$ is satisfied,  fall on a rather smooth curve 
about which $\theta$ nutates.  
At these  inflection points of $\theta$, the right-hand side (RHS) of Eq.(\ref{EulerYY}) vanishes. 
Here we note that,  unless $\sin \theta \approx 0$,  the second term $-h(\theta)F_X$  may be neglected 
as compared with the first term,  since Fig.\ref{TraVelocity} shows the smallness of $v_{PX}$.  
Then solving $V(\xi, \theta, J)=0$ for $\xi$, we obtain
\begin{equation}
\xi_{\pm}=\frac{J\pm \sqrt{J^2+4AMg (R \cos \theta-a)(R-a\cos \theta)a}}
{2(R\cos\theta-a)}~. \label{xiSolution}
\end{equation}
We expect that at the inflection points of $\theta$ and if not  $\sin \theta \approx 0$, 
$\xi$ takes the values which are  either on the curve $\xi_+$ or $\xi_-$. 
In the limit $(AMgR^2a)/J^2 \ll 1$ and $a \ll R$, which is true in these simulations, we have 
\begin{equation}
\xi_+\approx\frac{J}{R\cos\theta-a},  \qquad
\xi_-\approx-\frac{AMgaR}{J} \label{xiSolutionMinus}~,  
\end{equation}
and thus $\xi_- / \xi_0 \approx 0$. 

Fig.{\ref{BehaviorXi(a)}} shows the result of the
simulation with an initial value $\theta_0= 0.1$ rad.
The variable $\xi$, starting from  a large positive 
value $\xi_0=Cn_0$, begins  to fluctuate around the curve $\xi_+$.  
The fluctuation of $\xi$ becomes larger but $\xi$ is still positive for a while. 
The inclination angle $\theta$ decreases rapidly from the initial value $\theta_0$ and then starts to 
nutate. The amplitude of nutation becomes larger and the minimum value of $\theta$ 
decreases further.  And at a certain point where $\theta\approx 0$, the fluctuation of $\xi$ 
becomes so large  that $\xi$  takes negative values.  Then $\xi$  
starts to  fluctuate around the curve $\xi_-$ and $\theta$ is going to increase while nutating.  
The fluctuation of $\xi$ is getting smaller as $\theta$ is increasing,  but it  becomes 
large again when $\theta$ approaches $\pi$. We have observed in Fig.{\ref{BehaviorXi(a)}} that the
fluctuation of $\xi$ around the curve $\xi_+$ at the beginning soon shifts to the one 
around the curve $\xi_-$. For this rapid transition of $\xi$, the simulation shows that the system 
should pass through the phase where $\theta \approx 0$. 
When a simulation starts with a very small initial value $\theta_0$  as in Fig.{\ref{BehaviorXi(b)}}, 
then $\xi$  quickly moves to a fluctuation around the curve $\xi_-$. 

Let us look more closely the behavior of $\xi$ in Fig.{\ref{BehaviorXi(a)}} at an early stage 
(to be specific, between $0\!<\!t\!<\!1$ sec).  Recall $J=\xi_0(R\cos \theta_0-a)$. 
Then Eq.(\ref{EulerXXX}) gives
$U(\xi, \theta, J)|_{t=0}=-\xi_0\cos \theta_0$, which is  large and negative. At the very beginning of time, 
Fig.\ref{TraVelocity} (a) shows that 
$v_{PY}/|\vv_P| \approx 1$,  and  thus we have $F_Y\approx -\mu Mg$. Also 
the term $\dot\xi\sin\theta$ on the 
left-hand side (LHS) of  Eq.(\ref{EulerXX}) may be neglected in the leading order as compared with the $h(\theta)F_Y$ 
term,  since $(d(\xi/\xi_0)/dt)\sin\theta\sim \sin\theta \times (1/{\rm sec})$ and $\sin\theta$ is  
small,  while $\mu Mg R/\xi_0\sim 2\times (1/{\rm sec})$.
Hence we find from (\ref{EulerXX}), 
\begin{equation}
\dot\theta \approx \frac{\mu Mg h(\theta)}{U(\xi, \theta,J)}<0~, \qquad {\rm at\ the\ very\ beginning,}
\end{equation}
which explains a rapid decrease of $\theta$ from an initial value $\theta_0$.

Along with the rapid decrease of $\theta$, Fig.\ref{TraVelocity} (a) shows that 
the slip velocity $\vv_P$ of the contact point $P$ tends to vanish. Then, in this region 
where $\theta$ is small and $\vv_P\approx \zerov$, the term $h(\theta)F_Y$ 
of the RHS of (\ref{EulerXX}) may be neglected while $U(\xi,\theta,J)$ is
expressed as $U(\xi,\theta,J)\!\approx\! (\xi_0-2\xi)$. Hence  
Eq.(\ref{EulerXX}) is reduced to
\begin{equation}
{\dot \xi} \sin\theta=(\xi_0-2\xi){\dot \theta}~, \label{Nutation}
\end{equation}
and its solution is given by
\begin{equation}
|2\xi-\xi_0|={\rm const.}\times\frac{1+\cos\theta}{1-\cos\theta}~.
\end{equation}
We observe in the simulation shown in Fig.\ref{BehaviorXi(a)} that the behavior of $\xi$ 
during the time $0.05 <t<0.4$ sec is approximately described as 
\begin{equation}
\frac{\xi}{\xi_0} \approx \frac{1}{2}+C_1 \frac{1+\cos\theta}{1-\cos\theta}, \label{XiTheta}
\end{equation}
with a positive constant $C_1$. Although the term $h(\theta)F_Y$ 
has been neglected to derive (\ref{Nutation}), the small effect of the frictional force still remains 
and it produces the nutation of $\theta$, which in turn gives $\xi$ a fluctuating behavior  around the  curve 
$\xi_+$  according to (\ref{XiTheta}). Along with nutation, the minimum value of $\theta$ 
further decreases and so the fluctuation of $\xi$ is getting  larger. 

Then at a certain point (at $t\approx 0.45$ sec),  
the behavior of $\xi$ shifts to the one which is, later on up to 1 sec, roughly described as
\begin{equation}
\frac{\xi}{\xi_0} \approx \frac{1}{2}-C_2 \frac{1+\cos\theta}{1-\cos\theta}, \label{XiTheta-}
\end{equation}
with a  positive constant $C_2$, and $\xi$ may take negative values. 
Actually $\xi$ fluctuates rapidly between positive and negative values. 
Also, with the shift of the behavior of $\xi$,  $U(\xi,\theta,J)$ turns to  always take   positive values.  
In this region,  $v_{PY}$ is small but positive  on the average  in time (see Fig.\ref{TraVelocity} (a)). 
Now taking the time average of both sides of (\ref{EulerXX}), we see $\overline{\dot \theta}$, 
the time average of  $\dot \theta$, is positive, since the LHS, $\overline{{\dot \xi}\sin\theta}$,  may be 
neglected while $\overline {h(\theta)F_Y}$ is negative.
Thus, from 0.45 sec to 1 sec, $\theta$ gradually increases while nutating.  
As $\theta$ is increasing, the effect of $\sin \theta$  on $\dot\xi$
in the LHS of (\ref{EulerXX})  gets weaker and the fluctuation of
$\xi$ becomes  smaller. In the end $\xi$  oscillates mildly about  a  negative 
value $\xi_{-}$.

When we start simulation with a very small initial value $\theta_0= 0.01$ rad as in Fig.{\ref{BehaviorXi(b)}}, 
$\xi$  quickly takes negative values and $\theta$ starts to increase. 
The fluctuations of $\xi$ and $\theta$ are much smaller than those in
Fig.{\ref{BehaviorXi(a)}}.  With a smaller $\theta_0$,  the center of mass $O$ receives 
less recoil  from the frictional force $\Fv$, which explains the smaller fluctuations 
for   $\xi$  and $\theta$.  

So far we have shown the result of the numerical analysis for a tippe top 
which belongs to Group II.  
Given a sufficiently large spin, the GBC for this tippe top, which is not fulfilled 
initially, will  soon be satisfied approximately, and the body will start to turn over. 
Actually the GBC, $\xi\!=\!0$, is modified to $\xi\!=\!\xi_{\rm m}\!\equiv\! -AMgaR/J$ 
and $|\xi_{\rm m}/\xi_0|\ll 1$.  This modification  has an only effect
of shifting the positions of $\theta_c$ (\ref{ThetaCritical}) and 
$\theta_f$ (\ref{ThetaFixed}) slightly.

\begin{figure}
  \begin{center}
    \includegraphics[width=0.7\textwidth]{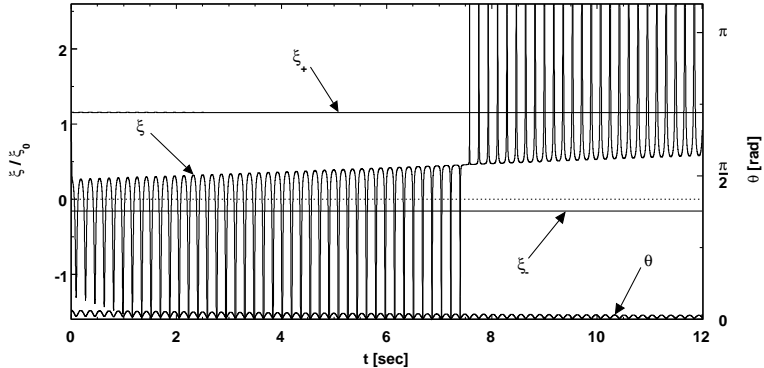}
    \caption{
      \label{GroupII-slow}
      Time evolution up to 12 sec of the variable $\xi$ and  the inclination angle $\theta$ for a 
tippe top of Group II with an initial 
      spin $n_0=$30  rad/sec. Input parameters and other initial conditions are the same as 
      in Figure \ref{BehaviorXi(a)}.
      The curves $\xi_{\pm}$ are given by 
      Eq.(\ref{xiSolution}).
    }
  \end{center}
\end{figure}

Empirically we know that when a given spin is not fast enough, the tippe top does 
not turn over and stays spinning with  its stem up. Later in Sec. 4, we argue that 
there exists a critical value for the initial spin given to the tippe tops of Group II and III.  
If the initial spin is below this critical value, then  even the tippe tops of Group II and 
III do not turn over. We have performed similar simulations as those in  
Figs. {\ref{BehaviorXi(a)}} and {\ref{BehaviorXi(b)}} with the same tippe top and 
the same initial conditions, except that the initial spins are below the critical value.
In these simulations we find that $\xi$, starting from a positive $\xi_0$,  first
fluctuates around the negative value $\xi_-$ and
then  returns to  positive values and  fluctuates around $\xi_+$, while 
the inclination angle $\theta$ remains  approximately zero. 
A typical example is shown in Fig. {\ref{GroupII-slow}}, where input parameters and 
initial conditions are the same as in Fig. \ref{BehaviorXi(a)} 
(and thus the tippe top for this simulation belongs to Group II), except that the initial 
spin $n_0$ is  30  rad/sec.  The critical value for the initial spin  is given by $n_1$ in (\ref{zeroStability4}) 
below and we have $n_1=$36 rad/sec for this case. 

\begin{figure}
  \begin{center}
    \begin{tabular}{cc}
 \includegraphics[width=0.49\textwidth]{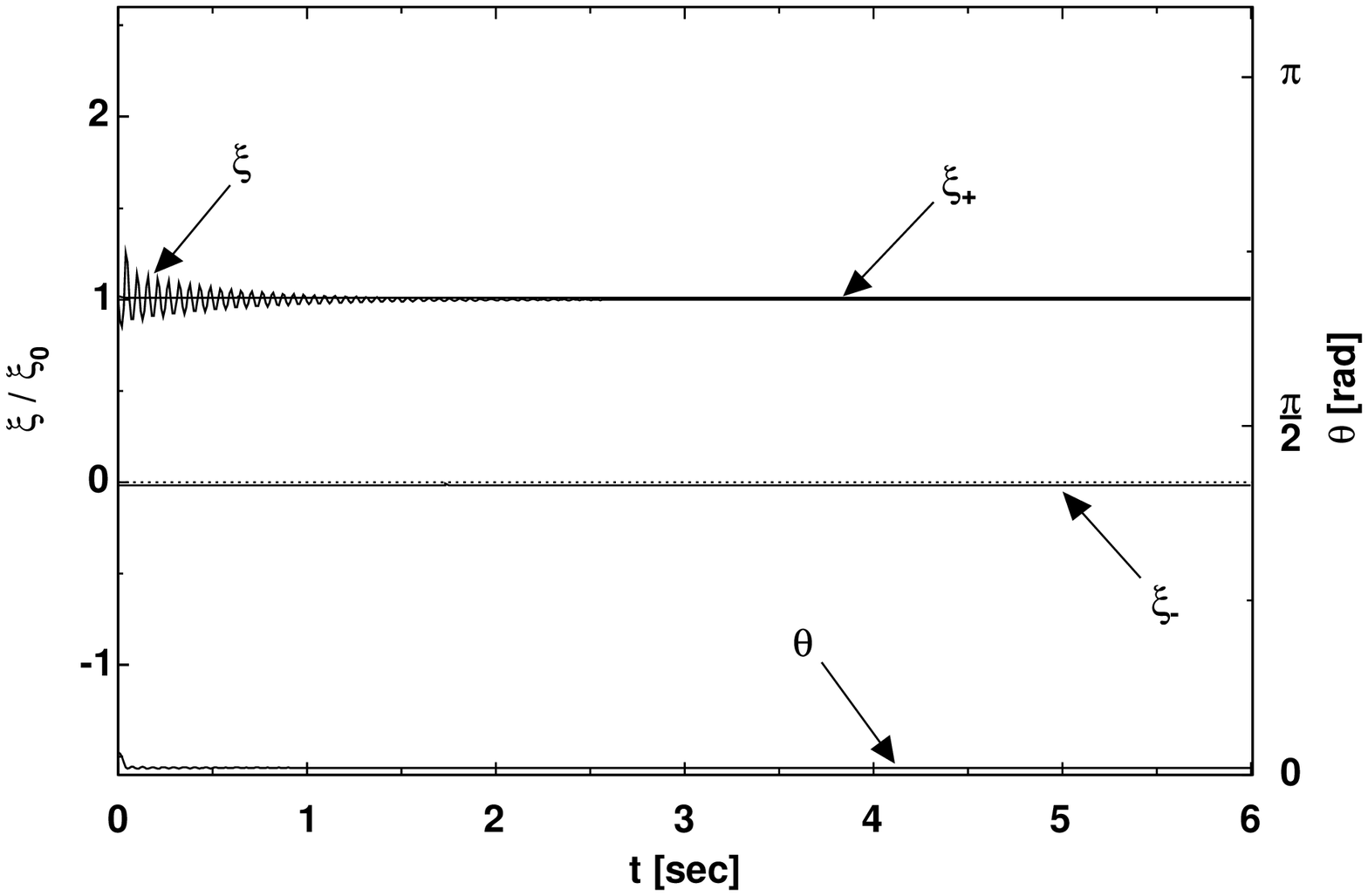} &
      \includegraphics[width=0.49\textwidth]{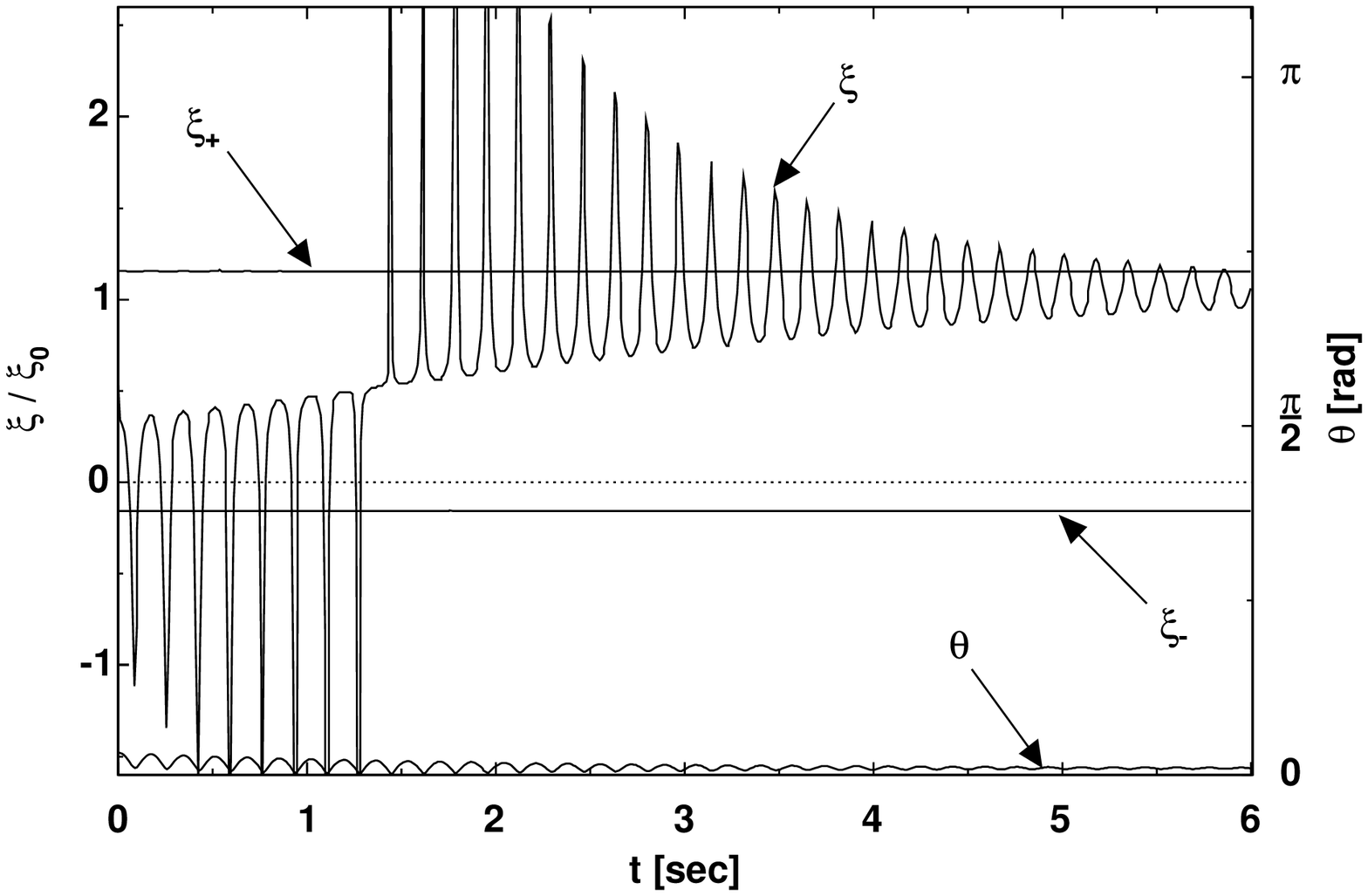} \\
      (a) & (b)
   \end{tabular}
    \caption{
      \label{GroupI-xi}
       Time evolution of the variable $\xi$ and the inclination angle $\theta$
       for a tippe top of Group I with  initial spins (a) $n_0=100$ rad/sec and 
(b) $n_0=30$ rad/sec.  Input parameters are 
      $R\!=\!1.5$ cm, $a\!=\!0.15$ cm (and thus  $a/R\!=\!0.1$), $M\!=\!15$ g, $g\!=\!980$ cm/sec${}^2$, 
      $A/C\!=\!0.85$, $C\!=\!2/5 M R^2$,
      $\mu\!=\!0.1$,
      $\Lambda\!=\!1$ cm/sec.   Other initial conditions are $\theta_0\!=\!0.1$ rad, 
      $\dot{\theta}_0\!=\!\Omega_0\!=\!0$,
      $\uv_0\!=\!\zerov$. The curves $\xi_{\pm}$ are given by 
      Eq.(\ref{xiSolution}).
    }
  \end{center}
\end{figure}

We also performed  simulations for the tippe tops of Group I, 
which are predicted to stay spinning at $\theta \approx 0$  however large a spin 
they are given.  
Plotted in Fig. {\ref{GroupI-xi}} are the time evolution of $\xi$ 
and  $\theta$ for 
a tippe top belonging to  Group I with  initial spins (a) $n_0=100$ rad/sec and 
(b) $n_0=30$ rad/sec. 
Given a large initial spin (Fig. {\ref{GroupI-xi}}(a)),  $\xi$ for a tippe top of Group I 
stays positive and  takes values very close to $\xi_+$. But, with a small initial spin 
(Fig. {\ref{GroupI-xi}}(b)),  $\xi$ changes from a 
positive $\xi_0$  to negative  and fluctuates around  $\xi_-$ for a while,  and then back to 
positive values again.  In both cases the tippe top stays spinning  at $\theta \approx 0$.

\begin{figure}
  \begin{center}
    \begin{tabular}{cc}
 \includegraphics[width=0.49\textwidth]{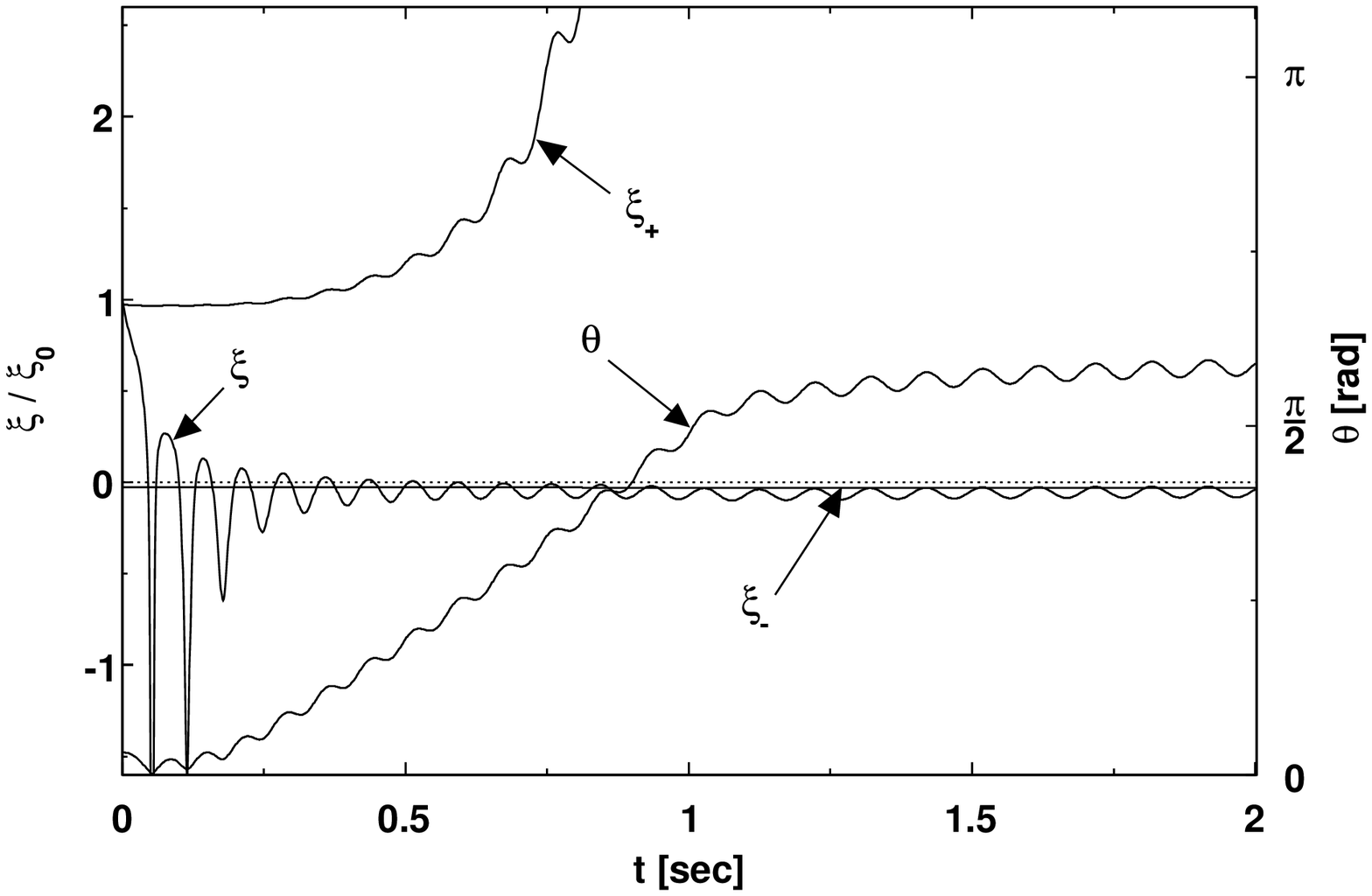} &
      \includegraphics[width=0.49\textwidth]{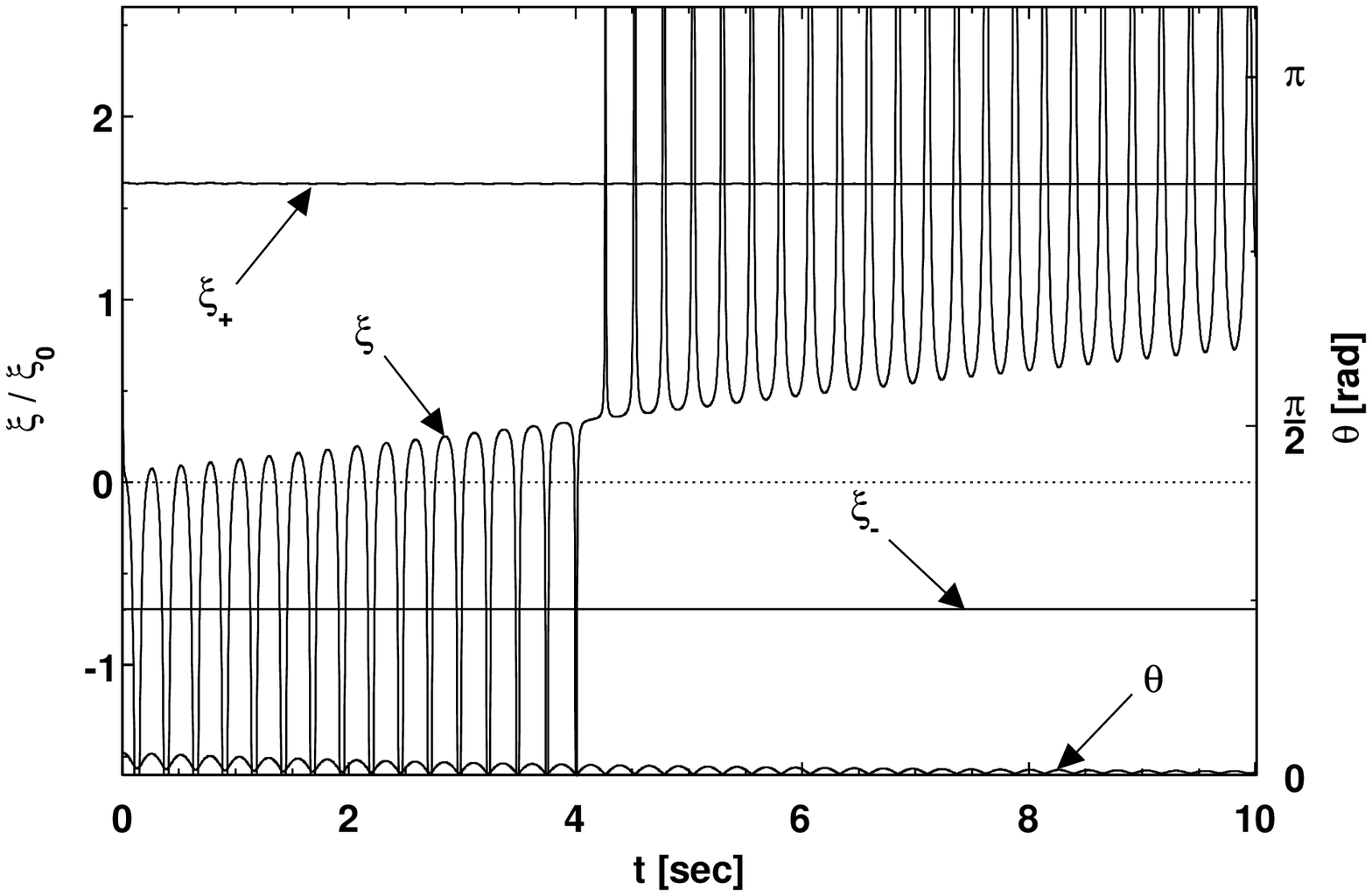} \\
      (a) & (b)
   \end{tabular}
    \caption{
      \label{GroupIII-xi}
       Time evolution of the variable $\xi$ and the inclination angle $\theta$
       for a tippe top of Group III with  initial spins (a) $n_0=100$ rad/sec and 
(b) $n_0=15$ rad/sec.  Input parameters are 
      $R\!=\!1.5$ cm, $a\!=\!0.225$ cm (and thus  $a/R\!=\!0.15$), $M\!=\!15$ g, $g\!=\!980$ cm/sec${}^2$, 
      $A/C\!=\!1.25$, $C\!=\!0.8\times (2/5) M R^2$,
      $\mu\!=\!0.1$,
      $\Lambda\!=\!1$ cm/sec.   Other initial conditions are $\theta_0\!=\!0.1$ rad, 
      $\dot{\theta}_0\!=\!\Omega_0\!=\!0$,
      $\uv_0\!=\!\zerov$. The curves $\xi_{\pm}$ are given by 
      Eq.(\ref{xiSolution}).
    }
  \end{center}
\end{figure}

For completeness we show, in Fig. {\ref{GroupIII-xi}},  typical examples of  
the time evolution of $\xi$ and  $\theta$ for 
a tippe top belonging to  Group III with  initial spins, (a) $n_0=100$ rad/sec and 
(b) $n_0=15$ rad/sec. 
Input parameters and other initial conditions are explained in the  caption of Fig. {\ref{GroupIII-xi}}.
The critical value for the initial spin  for the tippe top of Group III is  
given again by $n_1$ in (\ref{zeroStability4}) 
below and we have $n_1=$23.5 rad/sec for this simulation. When the initial spin $n_0$ is 
larger than the critical value $n_1$ (Fig. {\ref{GroupIII-xi}}(a)), the variable $\xi$ for a tippe top of Group III
shows a similar  behavior as the one presented in Fig. {\ref{BehaviorXi(a)}} for the tippe top of Group II.
To be specific, $\xi$ becomes small and fluctuates around the curve $\xi_-$ while $\theta$ increases. 
Note that the tippe top of Group III never turns over to the inverted position, $\theta\!=\!\pi$.
In the simulation of Fig. {\ref{GroupIII-xi}}(a), $\theta$ goes up to  the asymptotic angle 
$\theta_{\rm asymp}$, which is below the fixed point $\theta_f$(=2.21 rad.) derived from 
(\ref{ThetaFixed}). (See also the discussion on the plot in Fig. {\ref{ThetaFGroupIII}}).  
Given a smaller initial spin than $n_1$ (Fig. {\ref{GroupIII-xi}}(b)), $\xi$ for a tippe top of Group III,  
starting from a positive $\xi_0$,  
fluctuates around the negative value $\xi_-$ and
then  becomes positive and  fluctuates around $\xi_+$, while 
$\theta$ remains  approximately zero, a similar behavior as the one shown in Fig. 
\ref{GroupII-slow} for the case of a tippe top of Group II with an insufficient initial spin.

From these numerical analyses we see that the behavior of  $\xi$ is closely related to 
the inversion phenomenon of the tippe top. 
As 
the tippe top turns over, simulation shows that 
$\xi$ becomes small (in the sense $|\xi/\xi_0|\approx |\xi(R-a)/J|\ll1$) and takes 
values close to $\xi_-\approx \xi_m$, which implies that  the relation (\ref{GBCJellett}) is approximately 
satisfied. Conversely, when  the relation (\ref{GBCJellett}) holds,  it means that the center of mass of 
the tippe top goes up as $L_Z$ decreases.

\vspace{1cm}
\section{Stability and critical spin}
\subsection{Steady states} 

In Sec.3.2 we have studied the behaviors of the spinning tippe top when  
the gyroscopic balance condition $\xi\!=\!0$ is exactly satisfied.  The situation corresponds 
to the one in which the tippe top is given an infinitely large initial spin. 
Actually the initial spin given to the tippe top is finite and 
we know empirically that a tippe top with a small spin is stable and does not turn over. 
We will now consider how large an initial spin should be  for the tippe top to turn over. 
For that purpose we will study the steady states  of the tippe top and 
examine their stability.

Actually the steady states (or the asymptotic states) of the tippe top and their stability were
analyzed by Ebenfeld and Scheck [ES]~\cite{EbenfeldScheck}.  They used the total energy of the 
spinning top as a Liapunov function. Then the steady states were found as solutions 
of constant energy.  
%
%
The stability or instability of these states was determined by examining 
whether the Liapunov function assumes a minimum or  a maximum  at these states under 
the constraint of  Jellett's  constant.
%
The tippe top inversion was also analyzed  recently by Bou-Rabee, Marsden and Romero
[BMR]~\cite{BMR}  as a dissipation-induced  instability.  
BMR used the modified Maxwell-Bloch equations and an energy-momentum argument 
to determine the stability of the non-inverted and inverted states of the tippe top.

Here we take a different approach to this problem. 
And we discuss 
%
%
the stability of the steady states 
in terms of the initial spin velocity $n$ given at the non-inverted position $\theta \!=\!0$. 
Recently,  Moffatt, Shimomura and Branicki [MSB] made 
a linear stability analysis of the spinning motion of  spheroids~\cite{MSB}. They identified 
the steady states,  and then discussed 
%
%
their stability
and found the critical angular velocity needed for the rise of the body. 
In order to find the steady states of the spinning tippe top, we adopt the  method taken by 
MSB for the case of  spheroids.  But for the  stability analysis of  the steady states, we develop a new 
stability criterion  which is  different from the ones used by ES, BMR and MSB. 

Our approach to the stability problem of the tippe top is as follows. 
Once a steady state is known, the system is perturbed around the steady state. 
Particularly we focus our attention  on the 
variable $\theta$, which is perturbed to
\begin{equation}
\theta=\theta_s+\delta\theta~, 
\end{equation} 
where $\theta_s$ is a value at the steady state and   $\delta\theta$ is a small quantity. 
Using the equations of motion, we obtain,  under the linear approximation,
a  first-order ODE for $\delta\theta$ of the following form:   
\begin{equation}
\delta {\dot \theta}=H\Bigl(n_s, \Omega_s, \theta_s, \frac{A}{C}, \frac{a}{R}\Bigr)\delta\theta~, 
\label{EqForStability}
\end{equation}
where $n_s$ and $\Omega_s$ are  values taken at the steady state. 
Equation (\ref{EqForStability}) implies that the change of $\delta\theta$ is 
governed by the sign of the function $H$. If $H$ is positive (negative),  $|\delta\theta|$ will 
increase (decrease) with time. Therefore, we conclude that 
{\it  when  $H$ is negative (positive), then the state is stable (unstable)}.
This is the criterion for  stability of the steady state,  which we will  use in this paper.  
The stability criterion in this work is derived from an intuitive analysis of the equations of motion. 
We check in Appendices C and D that they are consistent with those derived by ES and BMR 
which are based on   mathematically rigorous methods.

\bigskip

Superficially the above criterion (\ref{EqForStability}) seems quite different from the one used by
ES~\cite{EbenfeldScheck},  but actually  we have found that both are equivalent and, therefore, 
our results are consistent with theirs.  In Appendix C we will show   the equivalence of both 
criteria and that the stability conditions of the steady states which we will obtain coincide with 
the ones found by ES. 
After all, ES utilized the total
energy (an integral form)~\cite{EbenfeldScheck}, while we will use equations of motion (differential forms).

The criterion  (\ref{EqForStability})  for the stability of the tippe top also seems different from the ones used by 
BMR~\cite{BMR}, which were derived from the tippe top modified Maxwell-Bloch 
equations. In order to obtain the stability criteria, both BMR and we linearize equations of motion  
about the steady states and use sliding friction, which is assumed to be an analytic function 
of the slip velocity, as the main mechanism behind tippe top inversion. 
Thus it is well expected that both criteria lead to 
the consistent results on the stability of the non-inverted and inverted states.
(The stability of the intermediate states have not been analyzed yet by means of
the modified Maxwell-Bloch equations). 
In Appendix D we will show that the expressions of the criteria provided in BMR become more transparent 
when they are    rewritten  in terms of the parameters and classification criteria used in this paper,  
and that they lead to the same  stability conditions  for the 
vertical spinning states which will be obtained later 
by using  the criterion  (\ref{EqForStability}). Besides, although BMR did not mentioned, 
the classification of tippe tops into three groups, Group I, II, and III, 
 is shown to be possible through the close examination of  the criteria in BMR.

\bigskip

The steady states of the spinning motion of the tippe top are obtained from the equations of motion 
(\ref{EulerOmega}-\ref{Eulern}) and (\ref{EqCMX}-\ref{EqCMZ}) by setting 
${\dot\Omega}\!=\!{\ddot\theta}\!=\!{\dot\theta}\!=\!{\dot n}\!=\!{\dot u_{OX}}\!=\!{\dot
u_{OY}}\!=\!{\dot u_{OZ}}\!=\!0$~\cite{MSB}.  Since we assume that the sliding 
friction (\ref{Friction}), i.e., a modified version of Coulomb law\footnote{Recall that the exact 
Coulomb friction (\ref{Coulomb}) is non-analytic at  $\vv_P=\zerov$ and  a 
nonlinear friction law that would not appear in the linear approximation~\cite{Or}.}, is the only frictional 
force present,  the
energy equation 
\begin{equation}
\frac{dE}{dt}=\Fv\cdot\vv_P=-\mu N \frac{\vv^2_P}{|\vv_P(\Lambda)|} \label{EnergyEq}
\end{equation}
shows that $\vv_P=\zerov$ and $\Fv\!=\!\zerov$ at the steady states~\cite{MSB}. 
Thus we obtain for the steady states  of the tippe top,
\begin{subequations}
\begin{eqnarray}
&&u_{OX}=0, \label{uOXzero}\\
&&\Omega u_{OY}=0, \\
&&\Omega(Cn-A\Omega  \cos\theta)\sin\theta+Mga\sin\theta=0, \\
&&u_{OY}+\left\{ R(n-\Omega  \cos\theta )+a \Omega\right\}\sin\theta=0, \label{vPYzero}
\end{eqnarray}
\end{subequations}
where $N\!=\!Mg$ and the velocity equations (\ref{vPX}-\ref{vPY}) and (\ref{VelP}) have been used. 
The solutions for Eqs.(\ref{uOXzero}-\ref{vPYzero}) are:

\bigskip

\noi
i) Vertical spin state at $\theta=0$~:
\begin{equation}
u_{OX}=u_{OY}=0~, \quad\theta=0~, \qquad n\ \ {\rm arbitrary}, \quad 
\Omega\ \ {\rm undefined}~, \label{SteadyThetaZero}
\end{equation}
which is a spinning state about the axis of symmetry 
with the center of mass  below the sphere' center.

\bigskip

\noi
ii) Vertical spin state at $\theta=\pi$~:
\begin{equation}
u_{OX}=u_{OY}=0~, \quad\theta=\pi~, \qquad n\ \ {\rm arbitrary}, \quad 
\Omega\ \ {\rm undefined}~,\label{SteadyThetaPi}
\end{equation}
which is an overturned  spinning state about the axis of symmetry 
with the center of mass  above the sphere' center. 

\bigskip 
\noi
iii) Intermediate states:
\begin{subequations}
\begin{eqnarray}
&&u_{OX}=u_{OY}=0~, \quad  0<\theta<\pi ~,\nn\\
&&\Omega(Cn-A\Omega \cos \theta)+a Mg=0~, \label{InterA}\\
&& R(n-\Omega  \cos\theta )+a \Omega=0~. \label{InterB}
\end{eqnarray}
\end{subequations}
The elimination of $n$ from (\ref{InterA}) and (\ref{InterB}) gives
\begin{equation}
\Omega^2=\frac{Mga}{(A-C)\cos\theta+C\frac{a}{R}}~.
\end{equation}
The necessary (but not sufficient) condition for the existence of such states is
\begin{equation}
\left( \frac{A}{C}-1 \right)\cos\theta+\frac{a}{R}>0~.
\end{equation}
Recall (\ref{VPzero}) which was used for the classification of tippe tops into three groups 
in Sec. 3.2. 
Thus, intermediate states  may exist at  
$\theta\!>\!\theta_c\!=\!\cos^{-1}\!\left(\frac{a}{R(1-\frac{A}{C})}\right)$ 
for the tippe top of Group I ~($\frac{A}{C}<1\!-\!\frac{a}{R}$), 
at $\theta$ between 0 and $\pi$ 
for Group II~($1\!-\!\frac{a}{R}<\frac{A}{C}<1\!+\!\frac{a}{R}$), 
and at $\theta<\theta_f\!=\!\cos^{-1}\!\left(\frac{a}{R(1-\frac{A}{C})}\right)$
for Group III~($1\!+\!\frac{a}{R}<\frac{A}{C}$). 

\bigskip

There appear, in total, three categories of steady states for a loaded sphere version of the tippe top.
%

\subsection{Stability analysis of the steady states}
The turnover of the tippe top is associated with the effect of the sliding friction 
(with a coefficient $\mu$) at the point of contact $P$.  Near the steady states, we know that
$\vv_P\approx\zerov$,  which is equivalent to the situation where $\mu \approx 0$.  Thus for the stability 
analysis of the steady states, we consider the limiting case of 
$\mu \ll 1$~\cite{MSB}. Since we expect
$\frac{d}{dt}\!\sim\!  {\cal O}(\mu)$ near the steady states, 
we have ${\dot \theta}\!\sim\! {\cal O}(\mu)$
and  ${\ddot \theta}\!\sim\! {\cal O}({\mu}^2)$. 
Eq.(\ref{vPY}) shows $v_{{\rm rot}PY}\!\sim\!  {\cal O}(1)$, which leads to $v_{PY}\!\sim\! {\cal O}(1)$  
and, hence,  $F_Y\!\sim\!  {\cal O}({\mu})$. Then (\ref{EqCMY})  gives $u_{OX}\!\sim\!  {\cal O}({\mu})$,
and $v_{PX}\!\sim\!  {\cal O}({\mu})$ from 
(\ref{vPX}) and (\ref{VelP}),   and thus we have $F_X\!\sim\!  {\cal O}({\mu}^2)$
 and  $u_{OY}\!\sim\!  {\cal O}({\mu}^2)$.

The above order estimation in $\mu$ near the steady states leads to 
the primary balance  in (\ref{Eulertheta}) which holds
at leading order in $\mu$~\cite{MSB},
\begin{equation}
\Omega(Cn-A\Omega \cos \theta)+a Mg=0~. \label{StabilityOmega}
\end{equation}
Note that with a sufficiently large $\Omega$, 
Eq.(\ref{StabilityOmega}) reduces to $\xi\!=\!Cn\!-\!A\Omega \cos \theta\!=\!0$, the GBC.

\subsubsection{Stability of the vertical spin state at $\theta=0$}

The angle $\theta$ is perturbed from $\theta\!=\!0$, and we take $\theta\!=\!\delta\theta \ll1$. 
In the linear approximation we may take $n={\rm const.}$,   since 
Eq.(\ref{Eulern}) implies that $\dot n$ is quadratic in small quantities 
(note $F_Y\!\sim\! {\cal O}({\mu})$).
With $\cos \delta\theta=1$, the primary balance  (\ref{StabilityOmega}) gives 
\begin{equation}
\Omega=\frac{1}{2A}\left\{ Cn\pm \sqrt{(Cn)^2+4AMga}  \right\}~. \label{Omegaat0}
\end{equation}
In this approximation $\Omega$ is also a constant.  Then Eq.(\ref{EulerOmega}) gives
\begin{equation}
\delta \dot \theta=\frac{R-a}{2A\Omega-Cn}{\mu}Mg \frac{v_{PY}}{\vert \vv_P (\Lambda) \vert}~,
\end{equation}
and we may take 
\begin{equation}
v_{PY}=\left\{ R(n-\Omega )+a \Omega\right\}\delta\theta, \label{vPYthetazero}
\end{equation}
since $u_{OY}\sim {\cal O}({\mu}^2)$. Hence, we require for the stability at  $\theta=0$
\begin{equation}
\frac{R(n-\Omega )+a \Omega}{2A\Omega-Cn}<0~. \label{zeroStability1}
\end{equation}
Using the expressions of both ``$+$'' and ``$-$'' solutions for $\Omega$  in (\ref{Omegaat0}), 
the above condition is rewritten as 
\begin{equation}
\pm\Bigl( \frac{2AR}{R-a}-C  \Bigr)n<\sqrt{(Cn)^2+4AMga}~,
\end{equation}
which gives 
\begin{equation}
n^2\Bigl\{ \frac{A}{C} -(1-\frac{a}{R})\Bigr\}<\frac{Mga}{C}\Bigl(1- \frac{a}{R} \Bigr)^2~.
\label{zeroStability2}
\end{equation}

It is easily seen that  the requirement (\ref{zeroStability2}) is always satisfied 
for any spin velocity $n$  by the tippe top of Group I ($\frac{A}{C}< 1- \frac{a}{R} )$. 
As for the tippe top of Group II or III with $\frac{A}{C}>\left( 1- \frac{a}{R} \right)$, 
the requirement (\ref{zeroStability2}) is rewritten as
\begin{equation}
n^2<\frac{Mga}{C\{ \frac{A}{C} -(1-\frac{a}{R})\}}\Bigl(1- \frac{a}{R} \Bigr)^2=n_1^2~.
 \label{zeroStability3}
\end{equation}

\bigskip

The stability of the vertical spin state at $\theta=0$ is summarized as follows: For the tippe top of Group I with
$\frac{A}{C}<\left( 1- \frac{a}{R} \right)$,  the spinning state at $\theta=0$ is stable for any spin $n$, 
while for the tippe top of Group II or III with $\frac{A}{C}>\left( 1- \frac{a}{R} \right)$ 
we require $n<n_1$ for its stability.  In other words, the tippe top of Group II or III  becomes 
unstable at $\theta=0$ if 
\begin{equation}
n(\theta=0)>n_1=\sqrt{\frac{Mga}{C\{ \frac{A}{C} -(1-\frac{a}{R})\}}}\Bigl(1- \frac{a}{R} \Bigr)
 \label{zeroStability4}
\end{equation}

\subsubsection{Stability of the vertical spin state at $\theta=\pi$}

A similar analysis can be made for the stability of the spinning state at $\theta=\pi$. 
Now put $\theta=\pi-\delta\theta'$ with $\delta\theta'\ll1$.
Again we may take $n={\rm const.}$,  but note that $n$ may be negative near  $\theta\!=\!\pi$.  
With $\cos\theta\!=\!-1$,  the primary balance  (\ref{StabilityOmega}) gives 
\begin{equation}
\Omega=\frac{1}{2A}\left\{ -Cn\pm \sqrt{(Cn)^2-4AMga}  \right\} \label{OmegaatPi}~.
\end{equation}
In order for $\Omega$ to have a real solution, we require 
\begin{equation}
|n|> \frac{2\sqrt{AMga}}{C}=n_2~.\label{StabilityOmegaPi}
\end{equation}
For $|n|<n_2$, the spin is insufficient to overcome the effect of gravity  and
the orientation becomes unstable~\cite{MSB}. 

With  $\dot \Omega =0$, Eq. (\ref{EulerOmega}) gives
\begin{equation}
\delta\dot \theta'=-\frac{R+a}{2A\Omega+Cn}{\mu}Mg \frac{v_{PY}}{\vert \vv_P (\Lambda) \vert}~,
\end{equation}
and we may take,
\begin{equation}
v_{PY}=\left\{ R(n+\Omega )+a \Omega\right\}\delta\theta'.    \label{vPYthetapi}
\end{equation}
Hence we require for the stability at  $\theta=\pi$,
\begin{equation}
\frac{R(n+\Omega )+a \Omega}{2A\Omega+Cn}>0 ~.\label{piStability1}
\end{equation}
Using the expressions of both ``$+$'' and ``$-$'' solutions for $\Omega$  in (\ref{OmegaatPi}), 
the above condition gives
\begin{equation}
n^2\Bigl\{ (1+\frac{a}{R})-\frac{A}{C} \Bigr\}>\frac{Mga}{C}\Bigl(1+ \frac{a}{R} \Bigr)^2~.\label{piStability2}
\end{equation}

First, the requirement (\ref{piStability2}) is never satisfied by the
tippe top of Group III  $(\frac{A}{C}\!>\!1\!+\!\frac{a}{R})$. So 
the tippe top of Group III is unstable at $\theta=\pi$. Actually it never turns over to the position with 
$\theta=\pi$.  For the tippe top of Group I or II which satisfies $\frac{A}{C} <(1\!+\!\frac{a}{R})$,
the requirement (\ref{piStability2}) becomes
\begin{equation}
n^2>\frac{Mga}{C\{ (1+\frac{a}{R})-\frac{A}{C} \}}\Bigl(1+ \frac{a}{R} \Bigr)^2=n_3^2~.
 \label{piStability3}
\end{equation}
Note that $n_3^2\ge n_2^2$. 

\bigskip

The stability of the vertical spin state at $\theta=\pi$ is summarized as follows: For the tippe top 
of Group III ($1+ \frac{a}{R} <\frac{A}{C}$), the spinning state at $\theta\!=\!\pi$ is unstable for any  spin
$n$,  while for the tippe top of Group I or II with 
$\frac{A}{C}<(1\!+\!\frac{a}{R})$,  the state at $\theta=\pi$ is stable if 
\begin{equation}
|n(\theta=\pi)|>\sqrt{\frac{Mga}{C\{ (1+\frac{a}{R})-\frac{A}{C} \}}}\Bigl(1+ \frac{a}{R} \Bigr)=n_3~.
 \label{piStability4}
\end{equation}

\subsubsection{Stability of the intermediate state}

We have learned in Sec.4.2.1 that the spinning state of Group I at $\theta\!=\!0$ 
is stable. We  also know from the discussion in Sec.4.1 that the intermediate steady states of Group I, 
if they exist, must occur at   
$\theta\!>\!\theta_c=\cos^{-1}\left(\frac{a}{R(1-\frac{A}{C})}\right)$. This implies that 
the spinning motion of Group I near $\theta\!=\!0$ does not shift to  
a possible intermediate steady state. 
On the other hand,  the tippe tops of Group II and III  become 
unstable at $\theta\!=\!0$ when they are spun with a sufficiently large initial spin
$n(\theta\!=\!0)>n_1$,  where $n_1$ is given by (\ref{zeroStability4}),  and they will start to turn over.
Here we are interested in the intermediate steady states of the tippe top which are 
reached from the initial spinning position near $\theta\!=\!0$. Therefore, in this subsection, we focus on the  
possible steady states only for the tippe tops of Group II and III, and examine their stability.

The Jellett's constant given by (\ref{JellettConstant}) or (\ref{JellettConstant2}) is rewritten 
as 
\begin{equation}
J=Cn(R\cos \theta-a)+A\Omega R \sin^2 \theta~. \label{JellettConstant3}
\end{equation}
Now Eqs.(\ref{InterA}) and (\ref{InterB}) and the above expression of J 
completely determine the  intermediate steady states. They are derived by solving
\begin{equation}
\kappa\Bigl[(\frac{A}{C}-1)\cos\theta+\frac{a}{R}\Bigr]=
\Bigl\{(\cos\theta-\frac{a}{R})^2+\frac{A}{C}\sin^2\theta\Bigr\}^2 ~,\label{EqForIntermediate}
\end{equation} 
where 
\begin{equation}
\kappa=\frac{J^2}{MgaCR^2}~.
\end{equation} 
Define the following function:
\begin{equation}
F(x)=\frac{f_2(x)}{f_1(x)}\label{EqForIntermediate2}~,
\end{equation}
where $x=\cos\theta$ and 
\begin{subequations}
\begin{eqnarray}
f_1(x)&=&(\frac{A}{C}-1)x+\frac{a}{R}~,\label{f1}\\
f_2(x)&=&\Bigl\{(x-\frac{a}{R})^2+\frac{A}{C}(1-x^2)\Bigr\}^2~.\label{f2}
\end{eqnarray}
\end{subequations}
Then, Eq.(\ref{EqForIntermediate}) is rewritten as 
\begin{equation}
F(x)=\kappa~.    \label{EqForIntermediate3}
\end{equation}
Since $f'_2(x)=-4\sqrt{f_2(x)}f_1(x)$, we obtain
\begin{subequations}
\begin{eqnarray}
F'(x)&=&-4\sqrt{f_2(x)}-\frac{f_2(x)}{[f_1(x)]^2}(\frac{A}{C}-1)~, \label{DerivG}\\
F''(x)
&=&\frac{2}{[f_1(x)]^3}\Bigl\{ \Bigl([f_1(x)]^2+(\frac{A}{C}-1) \sqrt{f_2(x)}  \Bigr) ^2
+3 [f_1(x)]^4 \Bigr\} >0~.  \label{DerivDerivG}
\end{eqnarray}
\end{subequations}
 
\noi
The condition for the initial spin $n(\theta\!=\!0)>n_1$ means $J>Cn_1(R-a)$.  
Using (\ref{zeroStability4}),  we find $\kappa>(1\!-\!\frac{a}{R})^4/(\frac{A}{C}\!-\!1\!+\!\frac{a}{R})$, 
which leads to 
$\kappa>F(1)$. So we are looking for  solutions of $F(x)=\kappa$ with $\kappa>F(1)$.

\bigskip

\noi
(i)\quad Group II \quad $(1\!-\!\frac{a}{R}<\frac{A}{C}<1\!+\!\frac{a}{R})$

\noi
When $1\le \frac{A}{C} < (1\!+\!\frac{a}{R})$~,  $F'(x)<0$ and $F(x)$ is a monotonically decreasing function 
for $-1\le x \le 1$~. Hence, there is one and only one solution of $F(x)=\kappa$  at $x_s$ 
between $-1$ and 1, provided $F(1)<\kappa<F(-1)$~. 
Otherwise, there is no solution, which
means that  there exists no intermediate steady state. 
Expressing $J$ with the initial spin 
at $\theta\!=\!0$  as $J=Cn(\theta\!=\!0)(R\!-\!a)$, we find that 
the condition $\kappa<F(-1)$ gives 
\begin{equation}
n(\theta=0)<\sqrt{\frac{Mga}{C\{ (1+\frac{a}{R})-\frac{A}{C} \}}}\frac{\Bigl(1+ \frac{a}{R} \Bigr)^2}
{\Bigl(1- \frac{a}{R} \Bigr)}=n_4~. \label{n4}
\end{equation}
Thus, in the case ~$1\le \frac{A}{C} \!<\!(1\!+\!\frac{a}{R})$, one intermediate steady state exists 
at $x_s$,  provided  that 
\begin{equation}
n_1<n(\theta=0)<n_4~.
\end{equation}

We know $F''(x)>0$  from (\ref{DerivDerivG}),  
 and so $F(x)$ is concave upward for $-1\le x \le 1$. 
If ~$(1\!-\!\frac{a}{R})< \frac{A}{C} < 1$~, then $F(x)$ may have a local minimum 
at a certain $x$ between $-1$ and 1. 
Recall that we are looking for the steady states which are  reached from the position near $\theta=0$ 
and that  the requirement for this is $\kappa>F(1)$. 
Hence, for the existence of such a steady state we need 
\begin{equation}
F(-1)>F(1)\qquad  {\rm and } \qquad F(-1)>\kappa>F(1)~. \label{CondsmallAC}
\end{equation}
The first condition $F(-1)>F(1)$ gives
\begin{equation}
\frac{A}{C}>1-\frac{a}{R}\frac{(1+\frac{a}{R})^4-(1-\frac{a}{R})^4}{(1+\frac{a}{R})^4+(1-\frac{a}{R})^4}
\equiv r_c~, \label{rc}
\end{equation}
and the second one $F(-1)>\kappa>F(1)$ leads to $n_1<n(\theta\!=\!0)<n_4$~.  Some tippe tops of Group II with 
$\frac{A}{C}<1$ satisfy  $F'(1)>0$ as well as the conditions (\ref{CondsmallAC}),  and thus $r_c<\frac{A}{C}<1$.  
For such tippe tops,  the corresponding $F(x)$ has a local minimum 
between  $x_d$ and 1,  where  $x_d$ is a solution of $F(x_d)=F(1)$. 
These tippe tops, therefore, have one  intermediate steady state at $x_s$ between $\!-\!1$ and   $x_d$ when 
the condition $n_1<n(\theta=0)<n_4$ is satisfied. 
See the discussion of case (c) in Fig.\ref{ThetaFGroupII}.
For the tippe tops of Group II with ~$(1\!-\!\frac{a}{R})< \frac{A}{C}< r_c$, there 
exists no intermediate state. We will see later, in the  discussion of case (d) in Fig.\ref{ThetaFGroupII},
that these tippe tops will turn over to $\theta=\pi$ once given a spin $n(\theta=0)>n_1$,   
since $F(-1)<F(1)$ and, hence, $n_1>n_4$ for these  tops.

\bigskip

\noi
(ii)\quad Group III \quad $(1\!+\!\frac{a}{R}<\frac{A}{C})$

\noi
Since $f_1(x)$ should be positive, the allowed region of $x$ is  $x_f<x\le 1$ with $x_f=\frac{a}{R(1-\frac{A}{C})}$~. 
Eq.(\ref{DerivG}) together with $(\frac{A}{C}-1)\!>\!0$ shows that $F(x)$ is a 
monotonically decreasing function  for $x_f<x\le 1$. Note that $F(x)$ positively 
diverges  when $x$ approaches $x_f$ from larger $x$. 
Hence, once $\kappa>F(1)$ is satisfied, $F(x)=\kappa$ has one and only one solution at $x_s$ such 
that $x_f<x_s<1$~.
In other words,  one intermediate steady state always exists  at $\theta_s(=\cos^{-1}x_s)$ 
between 0 and  $\theta_f(=\cos^{-1}x_f)$ for the tippe top of Group III,  if the condition  $n(\theta=0)>n_1$ 
is satisfied.  When $n(\theta=0)$ gets larger, the angle 
$\theta_s$ gets closer to $\theta_f$ but never crosses $\theta_f$.
In order for $\theta_s$ to reach $\theta_f$, $n(\theta=0)$ should be infinite.

\bigskip

Now we know that there exists an intermediate steady state for the tippe top of Group II 
with property $r_c\!<\!\frac{A}{C}\!<\!1\!+\!\frac{a}{R}$, when  
$n(\theta\!=\!0)$  satisfies  $n_1\!<\!n(\theta\!=\!0)\!<\!n_4$. 
Also there is an intermediate steady state for the tippe top of Group III with 
$(1\!+\!\frac{a}{R})<\frac{A}{C}$ if 
$n(\theta\!=\!0)\!>\!n_1$.  
Let ($n_s, \Omega_s,\theta_s$) represent such a steady state so that 
($n_s, \Omega_s,\theta_s$) are related by  Eqs.(\ref{InterA}) and (\ref{InterB}),  and 
suppose this state to be perturbed to
\begin{equation}
n=n_s+\delta n~, \quad \Omega=\Omega_s+\delta \Omega~,\quad \theta=\theta_s+\delta \theta~.
\end{equation}
Noting that $\dot\theta_s=0$ and $F_Y|_s=0$,  we find that the perturbed state satisfies 
\begin{equation}
\delta {\dot \theta}=-\frac{\mu Mg}{\Lambda}  \frac{R^2 \sin^2\theta_s}{C\Bigl\{ S^2_s
+ (\frac{A}{C} \sin\theta_s)^2\Bigr\}} D(x_s) \label{EqThetaStability2}
 \delta\theta~,
\end{equation}
where
\begin{equation}
D(x_s)=4\Bigl[f_1(x_s)\Bigr]^2+\Bigl(\frac{A}{C} -1  \Bigr)  \sqrt{f_2(x_s)} ~. \label{Dxs}
\end{equation}
The details of the derivation of (\ref{EqThetaStability2}) 
are given in Appendix B. 

If $\frac{A}{C}>1$, then $D(x_s)>0$.  Also when $r_c<\frac{A}{C}<1$, we find that 
$D(x_s)$ is still positive (see Appendix B). 
Thus  we observe from 
(\ref{EqThetaStability2}) that $\delta \dot \theta \propto \delta \theta$ with a negative 
constant at the intermediate steady state, which means that this state is indeed stable.

\bigskip

Finally it is emphasized that the spinning state of  the tippe top of Group I 
is stable at $\theta\!=\!0$ and the top will not turn over from the position near $\theta\!=\!0$.
On the other hand, the tippe top of Group III, when given a 
sufficiently large spin near the position $\theta\!=\!0$, will tend to turn over 
and approach the steady state at $\theta_s$ 
but never up to the inverted position at $\theta=\pi$. 

\bigskip

\subsection{Critical spin for inversion of the tippe top of Group II}

The tippe top of Group II will turn over to the inverted position at $\theta\!=\!\pi$ 
when it is given a sufficient initial spin. Let us estimate the critical value $n_c$ 
so that the spinning top with $n(\theta\!=\!0)>n_c$ reaches the inverted 
position.\footnote{ The idea is borrowed from Ref.\cite{MSB}, where  MSB
estimated the critical angular velocity above which a uniform prolate spheroid will rise to the 
vertical state under the assumption of the GBC and thus the existence of 
Jellett's constant.} 
Recall that Jellett's constant (\ref{JellettConstant3}) 
is invariant during the turnover from $\theta\!=\!0$ to $\theta\!=\!\pi$.
From the relation $Cn(\theta\!=\!0)(R\!-\!a)=Cn(\theta\!=\!\pi)(\!-\!R\!-\!a)$, 
we obtain
\begin{equation}
n(\theta=\pi)=-\frac{R-a}{R+a}n(\theta=0)~.
\end{equation}
We already know that we need $|n(\theta\!=\!\pi)|>n_3$
for the stability at $\theta\!=\!\pi$, where $n_3$ is given in (\ref{piStability4}).
Thus we find
\begin{equation}
n(\theta=0)>\sqrt{\frac{Mga}{C\{ (1+\frac{a}{R})-\frac{A}{C} \}}}\frac{\Bigl(1+ \frac{a}{R} \Bigr)^2}
{\Bigl(1- \frac{a}{R} \Bigr)}=n_4~.
 \label{critical1}
\end{equation}
Also from the instability condition of the tippe top of Group II at $\theta\!=\!0$, we need
$n(\theta\!=\!0)>n_1$, where $n_1$ is given by (\ref{zeroStability4}).
Hence the condition for the tippe top of Group II to turn over up to $\theta=\pi$ 
is that the initial spin $n(\theta\!=\!0)$ should be larger than both $n_4$ and $n_1$. 
In fact, we observe $n_4>n_1$ for the tippe top with $r_c<\frac{A}{C}<1\!+\! \frac{a}{R}$,  
while  $n_4<n_1$ for the tippe top with $1\!-\! \frac{a}{R}<\frac{A}{C}<r_c$, where 
$r_c$ is given by (\ref{rc}).
Therefore, we obtain
\begin{equation}
  n_c = \begin{cases}
  n_4,  & {\rm for} \quad r_c<\frac{A}{C}<1\!+\! \frac{a}{R}~,\\
n_1,  & {\rm for} \quad  1\!-\! \frac{a}{R}<\frac{A}{C}<r_c ~.
   \end{cases}
\end{equation}

\subsection{Numerical analysis}

We now study the time evolution of the inclination angle $\theta$ from a spinning position near 
$\theta\!=\!0$.  Simulations are made with various values of 
$\frac{A}{C}$ and $\frac{a}{R}$, changing the input parameters $A$ and $a$.
Other input parameters are the same as those given in (\ref{InputParameter}).
Initial conditions are $\theta_0=0.01$ rad,  $\dot{\theta}_0\!=\!\Omega_0\!=\!0$, and $\uv_0\!=\!\zerov$, 
and the  initial value of the spin velocity $n_0$ is varied. Since we have chosen 
a very small $\theta_0$, we may consider $n_0$ as $n(\theta\!=\!0)$.

\begin{figure}
  \begin{center}
    \includegraphics[width=0.7\hsize]{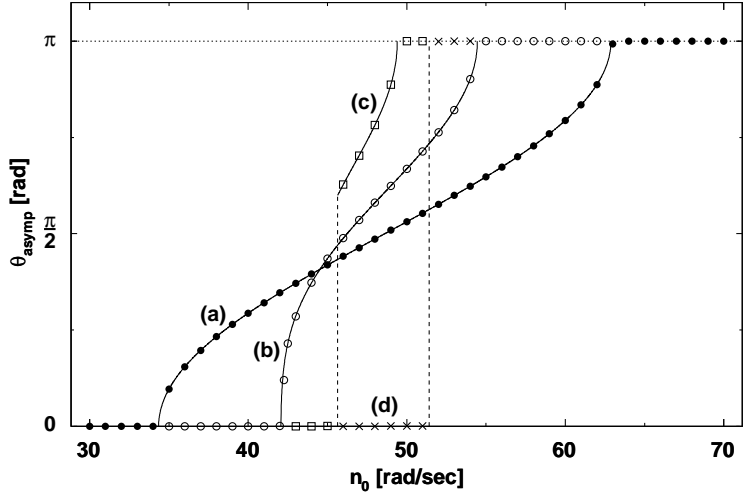}
    \caption{
      \label{ThetaFGroupII}
      The asymptotic value $\theta_{\rm asymp}$ as a function of the initial spin velocity $n_0$ 
      for tippe tops of Group II with various values of 
      $\frac{A}{C}$ and $\frac{a}{R}$;  (a) the one with $\frac{A}{C}=1$ and $\frac{a}{R}=0.15$; 
      the others have $\frac{A}{C}=0.95$ but different 
      $\frac{a}{R}$ such as  (b) $\frac{a}{R}=0.15$,  (c) $\frac{a}{R}=0.125$ and  (d) $\frac{a}{R}=0.1$. 
    }
  \end{center}
\end{figure}

Figure \ref{ThetaFGroupII} shows the asymptotic (final) angle of inclination, $\theta_{\rm asymp}$, as a function of 
$n_0$ for several  types of tippe tops of Group II with different values of 
$\frac{A}{C}$ and $\frac{a}{R}$;  (a) the one with $\frac{A}{C}=1$ and $\frac{a}{R}=0.15$; 
the others have $\frac{A}{C}=0.95$ but different 
$\frac{a}{R}$ such as  (b) $\frac{a}{R}=0.15$,  (c) $\frac{a}{R}=0.125$,  and  (d) $\frac{a}{R}=0.1$.   
The asymptotic angle $\theta_{\rm asymp}$ may be 0 or  $\pi$, or $\theta_s$,  the  
angle of  a possible intermediate steady state.

The symbols $\bullet$, $\circ$, $\diamond$ and $\times$ represent the results for the tippe tops
(a), (b), (c) and (d), respectively, and the thin solid curves (a), (b) and (c) are the 
trajectories obtained by solving  (\ref{EqForIntermediate}). We observe that the numerical results 
fall on the predicted curves. The values of $n_1(n_4)$,  in units of rad/sec, for 
the tops (a), (b), (c) and (d) are 34.4(62.9), 42.1(54.5), 45.7(49.4) and 51.4(44.4), respectively.  
In each case we see that the spinning state near
$\theta\!=\!0$  is stable when $n_0\!<\!n_1$. Once $n_0$ gets larger than $n_1$, 
the state becomes unstable and the tippe top turns over up to the
asymptotic angle $\theta_{\rm asymp}$.  For the tippe tops (a) and (b) 
the values of $\theta_{\rm asymp}$ grow with $n_0$ from 0 to $\pi$.   
On the other hand, 
the tippe top  (c) satisfies $r_c<\frac{A}{C}<1$ with $r_c=0.94$, 
and thus the intermediate steady state exists only at 
$\theta_s(=\theta_{\rm asymp})$ with $\theta_d\!<\!\theta_s\!<\!\pi$, where $\theta_d$ 
is a solution of  $F(\cos \theta_d)=F(1)$. We find $\theta_d=1.89$. 
Thus when $n_0$ gets larger than $n_1$ for the case of the tippe top (c), 
the asymptotic angle $\theta_{\rm asymp}$ jumps from 0 to $\theta_d$. 
When $n_0>n_4$,  $\theta_{\rm asymp}=\pi$ for the tops (a),  (b) and (c). 
In the case of the tippe top (d), we find $r_c=0.96$ and thus $\frac{A}{C}<r_c$, which leads to $n_1\!>\!n_4$. 
Therefore, there is no intermediate steady state, and the asymptotic angle $\theta_{\rm asymp}$ 
is 0 or $\pi$ depending on $n_0 \lessgtr n_1$. 

\begin{figure}
  \begin{center}
    \includegraphics[width=0.7\hsize]{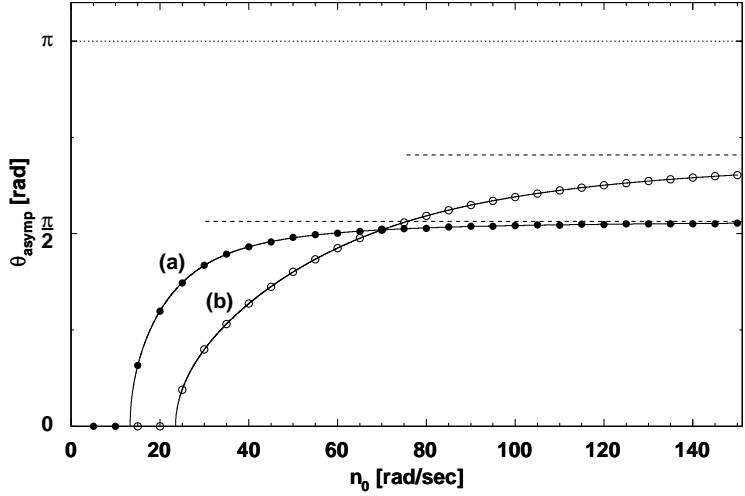}
    \caption{
      \label{ThetaFGroupIII}
      The asymptotic value $\theta_{\rm asymp}$ as a function of the initial spin velocity $n_0$ 
      for tippe tops of Group III:   (a) with  $\frac{A}{C}=1.25$ and $\frac{a}{R}=0.025$; and  (b) with
      $\frac{A}{C}=1.25$ and  $\frac{a}{R}=0.15$.
    }
  \end{center}
\end{figure}

We plot in Fig.\ref{ThetaFGroupIII}  the asymptotic angle $\theta_{\rm asymp}$ as a function of 
$n_0$ for the tippe tops of Group III;  (a)  with $\frac{A}{C}\!=\!1.25$ and 
$\frac{a}{R}\!=\!0.025$ and   (b) with $\frac{A}{C}\!=\!1.25$  and $\frac{a}{R}\!=\!0.15$.   
The symbols $\bullet$ and $\circ$ represent the results of simulation 
for the tippe tops (a) and  (b), respectively, and the thin solid curves (a) and (b) are the  
trajectories obtained by solving  (\ref{EqForIntermediate}).  We observe again that the
numerical results on  $\theta_{\rm asymp}$ for both tops (a) and (b) fall on the predicted curves. 
The values of $n_1$  for the tops  (a) and (b) are 13.3 and 23.5 rad/sec, respectively. In both cases 
the spinning position near $\theta=0$  is stable when $n_0$ is below $n_1$. Above $n_1$,  the value
of $\theta_{\rm asymp}$ grows with $n_0$ and approaches the fixed point $\theta_f$.  
The values of $\theta_f$ for the tops (a) and (b) are 1.67 and 2.21 rad, respectively.

For simulations we have used a modified version of the Coulomb friction $\Fv$ given in 
(\ref{Friction}).  The value $\theta_{\rm asymp}$ 
is not affected by the strength of the coefficient $\mu$.  
The strength of $\mu$ instead has an effect on the rate of rising of the tippe top.
If we use another form than (\ref{Friction}) for the sliding friction, 
and moreover,  it is expressed as a continuous function of  $\vv_P$ and vanishes at  
$\vv_P=\zerov$, then 
we still expect that we   get the same numerical results on  $\theta_{\rm asymp}$ {\it vs.}
$n_0$ as shown in Fig.\ref{ThetaFGroupII} and Fig.\ref{ThetaFGroupIII}.
This is due to the observation that
the numerical value  $\theta_{\rm asymp}$  has fallen on the predicted 
curves which are derived from (\ref{EqForIntermediate}) and that  
we have obtained  (\ref{EqForIntermediate}) using the property of
$\Fv$ which vanishes at  the steady states  together with $\vv_P$.

\begin{figure}
  \begin{center}
    \includegraphics[width=0.6\hsize]{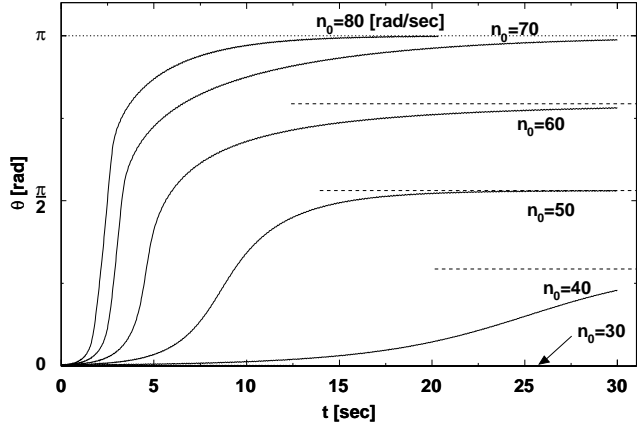}
    \caption{
      \label{ThetaEvolution}
      The time evolution of the angle $\theta$ for a tippe top of Group II from a spinning position near $\theta=0$. 
    }
  \end{center}
\end{figure}

Figure \ref{ThetaEvolution} shows the time evolution of the inclination angle 
$\theta$ for a tippe top of Group II from a spinning position near $\theta=0$ for various values of the initial spin velocity
$n_0$.  Input parameters and initial conditions are the same as before and we take $\frac{A}{C}=1$ and 
$\frac{a}{R}=0.15$. The asymptotic angles $\theta_{\rm asymp}$ which will be reached 
are 0, 0.92, 1.67, 2.49, $\pi$  and  $\pi$ rad for $n_0=$30, 40, 50, 60, 70 and 80 rad/sec, respectively.
Simulations with a modified version of the Coulomb friction (\ref{Friction}) show that the larger value of 
$n_0$ is given,  the faster the rate of rising becomes.

\vspace{1cm}

\section{Summary and Discussion}

We have examined an inversion phenomenon of the spinning tippe top, 
focusing our attention on its relevance to the gyroscopic balance condition (GBC), 
which was discovered by Moffatt and Shimomura in the study of the 
spinning motion of a hard-boiled egg. 
In order to analyze the GBC in detail for the case of the tippe top, we introduce a variable 
$\xi$ given by (\ref{xi}) so that 
$\xi\!=\!0$ corresponds to the GBC, and study the behavior of $\xi$. 
Contrary to the case of the spinning egg, the GBC is not satisfied initially for the tippe top.   
The simulation shows that, starting from a large positive value $\xi_0$, 
the variable $\xi$ for the tippe tops which rise, 
soon fluctuates around a negative but small value $\xi_m$  
such that $|\xi_m/\xi_0|\approx 0$. Thus for these tippe tops,  the GBC, though it is not fulfilled initially, 
will soon  be   satisfied approximately.   Once $\xi$ fluctuates around the value $\xi_{\rm m}$, 
these tops  become unstable and start to turn over.
On the other hand,  in the case of the tippe tops which 
do not turn over,  $\xi$ remains positive around $\xi_0$ or changes from  positive $\xi_0$ to 
negative values and then back to  positive values close to $\xi_0$ again. 

Under the GBC the governing equations for the tippe top are  much 
simplified  and, together with the geometry of the tippe top,  we obtain 
a first-order ODE for $\theta$ in the following form~\cite{MS}  
(see (\ref{EquForTheta}) or (\ref{EquTheta})) :
\begin{equation}
\frac{d\theta}{dt}=b(\theta)~. \label{EquThetaLast}
\end{equation}
It is noted that this equation has a remarkable resemblance to the renormalization group (RG) equation 
for the effective coupling constant $g$,
\begin{equation}
\frac{d g}{dt}=\beta(g)~, \label{RGEquation}
\end{equation}
which appears in quantum field theories for critical phenomena~\cite{WilsonKogut,JZJ} and 
high energy physics~\cite{Muta}. Here in (\ref{RGEquation}),  $t$ is expressed as $t={\rm ln}\lambda$ with a
dimensionless scale parameter $\lambda$. Provided  that $\beta(g)$ has a zero at $g=g_c$, we find that, 
if $\beta'(g_c)<0$, then $g(t)\rightarrow g_c$ as $t\rightarrow \infty$ ($\lambda\rightarrow \infty$), 
and while if $\beta'(g_c)>0$, $g(t)\rightarrow g_c$ as $t\rightarrow -\infty$ ($\lambda\rightarrow 0$). 
The limiting value $g_c$ of $g(t)$ is known as the ultraviolet (infrared) fixed point in the former 
(latter) case. Similarity between the two equations, (\ref{EquThetaLast}) and (\ref{RGEquation}),  
and the notion of the RG equation brought us to a consequence that tippe tops are classified into three groups, 
depending on the values of   $\frac{A}{C}$  and $\frac{a}{R}$. 
A resemblance of Eq.(\ref{EqForStability}) to the RG equation also gave us a hint that 
Eq.(\ref{EqForStability}) might serve as a criterion for stability of the steady state in Sec. 4.


The  criterion (4.2) is a 
first-order ODE for the (perturbed) inclination angle $\delta\theta$,  
and the results derived from this criterion 
coincide with those by ES and BMR which are obtained by  mathematically 
rigorous methods. The key ingredients in the process of arriving at this first-order ODE 
are the order estimation in $\mu$ near the steady states and an intuitive analysis of the equations 
of motion. The  criterion (4.2) can also be applied to the  stability analysis of 
other spinning objects. In fact we  have applied  (\ref{EqForStability}) to  the spinning motion of 
spheroids (prolate and oblate) which was recently examined in detail by MSB~\cite{MSB}, and 
we have obtained consistent results with theirs.

Finally we have assumed, in the present work,  a modified version of  Coulomb law   
(\ref{Friction}) for the sliding friction, since Coulomb friction (\ref{Coulomb}) 
is non-analytic and undefined at $\vv_P=0$. 
On the other hand, Cohen used Coulomb friction in his pioneering work on the tippe top~\cite{Cohen}, 
and analyzed its spinning motion numerically  for the first time.
He reported the result of a sample simulation in Fig.5 of his paper~\cite{Cohen}. 
The  Coulomb friction 
is realistic provided that $\vert \vv_P  \vert$ is away from zero, but its application to the spinning motion of 
the tippe top is very delicate. 
Near steady states  (i.e., near $\theta\!=\!0$ or $\pi$ or $\theta_s$), $\vv_P$ almost vanishes 
(see, for  example,   Fig.\ref{TraVelocity} (a) and (b)).  And there the $X$- and $Y$-components of
$\vv_P/|\vv_P|$ are changing signs rapidly and moreover non-analytically, and so are
the components of friction, $F_X$ and $F_Y$. Coulomb friction 
may not be adequate to be applied to such a situation.  In fact, 
Kane and Levinson~\cite{KaneLevinson} argued against the work of 
Cohen, because it did not include adequate provisions for transitions from sliding to rolling and 
vice versa. They reanalyzed the simulation of Cohen,  assuming Coulomb law for sliding friction, 
but also providing an algorithm  that rolling begins when $|\vv_P|\!<\! \epsilon$ 
(with $\epsilon\ll$ 1m/sec)   is satisfied,  together with  another algorithm for the transition 
from rolling to sliding.  They found that a transition from sliding to rolling occurs soon after the motion has
begun  and that values of $\theta$ remain below 0.077 rad  thereafter.  Or~\cite{Or} adopted a hybrid
friction law adding viscous friction, which is linearly related to $\vv_P$, to Coulomb friction.  
Other frictional forces such as the one which is due to pure rotation about 
the normal at the point of contact might have some effect.
After all it is safe to say that we have understood general features of the tippe top inversion.
But it would be not until we have had thorough knowledge of frictional force
that we completely understood the inversion phenomena of the tippe top.
{\it And yet, it flips over}.


\vspace{0.5cm}
\leftline{\large\bf Acknowledgments}
\vspace{0.5cm}
\noindent
We thank Tsuneo Uematsu for valuable information 
on the spinning egg and the tippe top.  We also thank  Yutaka Shimomura 
for introducing us to the paper~{\cite{MSB}} and 
for helpful discussions. 


\newpage
\appendix

\noindent
{\LARGE\bf Appendix}

\section{Equations of motion for the tippe top}

We enumerate the equations of motion which are used to analyze the spinning motion of the tippe top:
\begin{eqnarray}
A{\dot \Omega}\sin\theta&=&(Cn-2A\Omega\cos\theta){\dot \theta}
+(a-R\cos\theta)F_Y~,  \label{AEulerOmega}\\
A{\ddot \theta}&=&-\Omega(Cn-A\Omega\cos\theta)\sin\theta
-a\sin\theta N- h(\theta)F_X ~,  \label{AEulertheta}\\
C{\dot n}&=&R\sin\theta F_Y~.  \label{AEulern} \\
&&  \nn \\
M{\dot u}_{OX}&=&M\Omega u_{OY}+F_X ~, \label{AEqCMX}\\
M{\dot u}_{OY}&=&-M\Omega u_{OX}+F_Y ~,\label{AEqCMY}\\
M{\dot u}_{OZ}&=&N-Mg~. \label{AEqCMZ}\\
&&\nn\\
\Fv&=&-\mu N \frac{\vv_P}{|\vv_P(\Lambda)|}~, \ \  
{\rm with} \ \  \vert \vv_P (\Lambda) \vert=\sqrt{v^2_{PX}+v^2_{PY}+\Lambda^2}\\
&&\nn\\
v_{PX}&=&u_{OX}-h(\theta)\dot \theta~, \label{AvPX} \\
v_{PY}&=&u_{OY}+\left\{ R(n-\Omega  \cos\theta )+a \Omega\right\}
\sin\theta~, \label{AvPY}  \\
&&\nn\\
h(\theta)&=&R-a\cos\theta \\
u_{OZ}&=&a\sin\theta ~{\dot\theta}
\end{eqnarray}

\bigskip

\section{Stability  of the intermediate state}

There exists an intermediate steady state for the tippe top of Group II 
with property $r_c\!<\!\frac{A}{C}\!<\!1\!+\!\frac{a}{R}$, 
when an  initial spin 
$n(\theta\!=\!0)$  satisfies  $n_1<n(\theta\!=\!0)<n_4$. 
There is also an intermediate steady state for the tippe top of Group III if 
$n(\theta\!=\!0)>n_1$.  In this appendix we show that these steady states are stable. 

Near the steady states  the primary balance condition (\ref{StabilityOmega}) 
holds at  leading order in $\mu$. Differentiating both sides of (\ref{StabilityOmega})  with respect to $t$, 
we obtain
\begin{equation}
(Cn-2A\Omega \cos \theta){\dot \Omega}+C\Omega {\dot n}+A\Omega^2 \sin\theta {\dot\theta}=0~.
\end{equation}
Using (\ref{EulerOmega}) and (\ref{Eulern}), and eliminating $\dot \Omega$ and $\dot n$, we find
\begin{equation}
{\dot \theta}=\frac{-F_Y\Bigl\{ (a-R\cos\theta)(Cn-2A\Omega \cos \theta  )+A\Omega R
\sin^2\theta\Bigr\}}{(Cn-2A\Omega \cos \theta  )^2+(A\Omega \sin\theta)^2}~. \label{EqThetaStability}
\end{equation}
Let ($n_s, \Omega_s,\theta_s$) represent an intermediate steady state so that 
$n_s$, $\Omega_s$ and $\theta_s$  are related by (\ref{InterA}) and (\ref{InterB}),  and 
suppose this state to be perturbed to
\begin{equation}
n=n_s+\delta n~, \quad \Omega=\Omega_s+\delta \Omega~,\quad \theta=\theta_s+\delta \theta~.
\end{equation}
Since $\dot\theta_s=0$ and $F_Y|_s=0$,  the perturbed state satisfies
\begin{equation}
{\delta \dot \theta}=-\delta F_Y \frac{RT_s}{C\Omega_s\Bigl\{ S^2_s
+ (\frac{A}{C} \sin\theta_s)^2\Bigr\}} ~,\label{EqThetaStability1}
\end{equation}
where
\begin{eqnarray}
S_s&=&-\frac{Cn_s-2A\Omega_s\cos\theta_s}{C\Omega_s}=2\frac{A}{C}\cos\theta_s-
(\cos\theta_s-\frac{a}{R})~,  \label{SS} \\
T_s&=&S_s(\cos\theta_s-\frac{a}{R})+\frac{A}{C}\sin^2\theta_s ~.\label{TS}
\end{eqnarray}
At leading order in $\mu$, we have
$v_{PY}\!=\! v_{{\rm rot}PY}$  (recall $u_{OY}\!\sim\! {\cal O}({\mu}^2)$), and thus we obtain from (\ref{vPY}),  
\begin{eqnarray}
\delta F_Y&=&-\frac{{\mu}Mg}{\Lambda}  \delta v_{{\rm rot}PY} \nn\\
&=&-\frac{{\mu}Mg}{\Lambda}  R\sin\theta_s\Bigl\{ \delta n+(\frac{a}{R}-\cos\theta_s)\delta\Omega +
\Omega_s\sin\theta_s \delta\theta  \Bigr\} \label{deltaFY}
\end{eqnarray}

Now we expect that the perturbed state still satisfies  the primary balance condition (\ref{StabilityOmega}), 
since  $A{\ddot \theta}$ and $F_X$ are ${\cal O}({\mu}^2)$. Then a variation around the 
steady state gives
\begin{equation}
\delta n-S_s\delta \Omega+
\frac{A}{C}\Omega_s\sin\theta_s\delta\theta =0~. \label{VariA}
\end{equation}
where $S_s$ is given by (\ref{SS}).
Also taking a variation of Jellett' constant (\ref{JellettConstant3}) around the steady state 
(and then, of course, we have $\delta J=0$), we obtain 
\begin{equation}
(\cos\theta_s-\frac{a}{R})\delta n +\frac{A}{C}\sin^2 \theta_s\delta  \Omega
+S_s\Omega_s\sin\theta_s\delta\theta=0~. \label{VariB}
\end{equation}
From (\ref{VariA}) and (\ref{VariB}), $\delta n$ and $\delta\Omega$  are expressed in terms of
 $\delta\theta$ as
\begin{subequations}
\begin{eqnarray}
\delta n&=&-\frac{1}{T_s}\Bigl\{ S^2_s
+ (\frac{A}{C} \sin\theta_s)^2\Bigr\}  \Omega_s\sin\theta_s\delta\theta~, \label{deltaNs}\\
\delta\Omega&=&-\frac{1}{T_s}\Bigl\{ S_s- \frac{A}{C} (\cos\theta_s-\frac{a}{R})\Bigr\}
\Omega_s\sin\theta_s\delta\theta~. \label{deltaOmegas}
\end{eqnarray}
\end{subequations}
Inserting these expressions into (\ref{deltaFY}), and then we obtain from (\ref{EqThetaStability1}) 
\begin{equation}
\delta {\dot \theta}=-\frac{\mu Mg}{\Lambda}  \frac{R^2 \sin^2\theta_s}{C\Bigl\{ S^2_s
+ (\frac{A}{C} \sin\theta_s)^2\Bigr\}} D(x_s)
 \delta\theta~,
\end{equation}
where
\begin{equation}
D(x_s)=4\Bigl[f_1(x_s)\Bigr]^2+\Bigl(\frac{A}{C} -1  \Bigr)  \sqrt{f_2(x_s)} ~,
\end{equation}
and  $x_s=\cos \theta_s$,  and Eqs.(\ref{f1}) and (\ref{f2}) have been used. 

If $1<\frac{A}{C}$, then $D(x_s)>0$. 
Also when $r_c<\frac{A}{C}<1$,  
$D(x_s)$ is still positive, which is explained as follows: 
The expression of (\ref{DerivG}) shows that the function $D(x)$ is related to $F'(x)$ as 
\begin{equation}
F'(x)=-\frac{\sqrt{f_2(x)}}{[f_1(x)]^2}D(x)~. 
\end{equation}
When $r_c<\frac{A}{C}<1$, an intermediate steady state at $x=x_s$ exists 
provided $F(-1)>F(1)$ and $F(-1)>\kappa>F(1)$~. At that point $F'(x_s)$ is negative, and thus 
$D(x_s)$ is positive.

\bigskip
\newpage
\section{Equivalence between the criterion of ES~\cite{EbenfeldScheck} and  Eq.(\ref{EqForStability})}

Ebenfeld and Scheck~\cite{EbenfeldScheck} analyzed the stability  of the spinning tippe top using 
the total energy as a Liapunov function and gave the stability criteria for the steady states. 
We take a different approach to this stability problem. First the system is perturbed around the steady state. 
Then, using the equations of motion and   under the linear approximation,  we obtain 
a  first-order ODE for $\delta\theta$ of the  form given in (\ref{EqForStability}). 
We make use of this equation and give a different stability criterion. In this appendix 
we show that both approaches are equivalent and thus they lead to the same conclusions on the 
stability conditions of   the steady states.

ES wrote  the total energy of the spinning top as the sum of two terms 
(ES-(33))\footnote{From now on, we write the equation ($\star\star$) of Ref.\cite{EbenfeldScheck} as 
ES-($\star\star$). The Jellett constant $\lambda$ defined by ES is related to our $J$ as 
$\lambda=J/R$.
}
\begin{equation}
E=E^{(1)}(\eta_3,L_\parallel)+E^{(2)}(\hat{{\etav}},{\Lv}_\perp,\dot{\sv}_{1,2})~, 
\end{equation} 
the second of which contains all the terms that will vanish at the steady states,  while the 
first depends on $\eta_3\equiv \cos\theta$ and  Jellett's constant $J$. In terms of the parameters 
used in this paper,   
$E^{(1)}$ and $E^{(2)}$ are expressed as follows:
\begin{eqnarray}
E^{(1)}&=&\frac{J^2}{2AR^2 G(\eta_3)}+MgR(1-\frac{a}{R}\eta_3)~, \\
E^{(2)}&=&\frac{1}{2}M\left( u^2_{OX} +u^2_{OY}+u^2_{OZ}\right)+\frac{1}{2}A\dot{\theta}^2 \nn\\
&&+\frac{(1-\eta^2_3)G(\eta_3)}{2C(1-\frac{a}{R}\eta_3)^2}\biggl\{ \xi 
+\frac{J\Bigl[ \eta_3-\frac{C}{A}(\eta_3- \frac{a}{R} )\Bigr]}{RG(\eta_3)} \biggr\}^2~, 
\label{E2}
\end{eqnarray}
with
\begin{equation}
G(\eta_3)=1-\eta_3^2+\frac{C}{A}(\eta_3- \frac{a}{R})^2~, \qquad  \quad \eta_3= \cos\theta
\end{equation} 
Note that ES set $R\!=\!1$. The condition $dE^{(1)}(\theta)/d\theta\!=\!0$ together with 
$u_{OX}\!=\!u_{OY}\!=\!u_{OZ}\!=\!0$  leads to the three solutions of the steady states: 
(i) vertical spin state at $\theta\!=\!0$ (\ref{SteadyThetaZero}), (ii) 
vertical spin state at $\theta\!=\!\pi$ (\ref{SteadyThetaPi}), and (iii)
intermediate states (\ref{EqForIntermediate}), or equivalently,  (\ref{InterA}-\ref{InterB}). 
It is recalled that we have obtained these solutions starting from equations of motion. 
At these steady states $E^{(2)}$  vanishes. For intermediate steady states,
the factor $\{ \xi 
+J[ \eta_3-\frac{C}{A}(\eta_3- \frac{a}{R} )]/RG(\eta_3) \}$ in (\ref{E2}) reduces to zero,  
due to  (\ref{InterA}-\ref{InterB}) and  Jellett's  constant given in (\ref{JellettConstant3}). 

Now we show that the  criterion, Eq.(\ref{EqForStability}), for the stability of the steady states is 
equivalent to the one  derived by ES~\cite{EbenfeldScheck}. 
For the stability  analysis of the steady states, the order estimation in $\mu$ near the steady states
is important, which has been pointed out by MSB in their work on
the linear stability analysis of the spinning motion of  spheroids~\cite{MSB}. 
As explained at the beginning of Sec. 4.2, near the steady states we have $\frac{d}{dt}\!\sim\!  {\cal O}(\mu)$, 
$v_{PX}\!\sim\!{\cal O}({\mu})$, $v_{PY}\!\sim\!v_{{\rm rot}PY}\!\sim\!  {\cal O}(1)$, and 
$u_{OY}\!\sim\!  {\cal O}({\mu}^2)$. Since $E^{(2)}$ is already ${\cal O}({\mu})$ (recall that it vanishes 
at the steady states), we have $\frac{dE^{(2)}}{dt}\!\sim\! {\cal O}({\mu}^2)$, while 
$\frac{dE^{(1)}}{dt}\!\sim\! {\cal O}({\mu})$.  Thus near the steady states,  
the energy equation  (\ref{EnergyEq}) is written  at  leading order in $\mu$ as
\begin{equation}
\frac{dE^{(1)}}{dt}=\frac{dE^{(1)}}{d\theta}{\dot \theta}=-\mu Mg \frac{v_{{\rm rot}PY}^2}{|\vv_P(\Lambda)|}~.
\label{EnergyEqLeading}
\end{equation} 
Suppose the steady states to be perturbed to
$n\!=\!n_s\!+\!\delta n, ~\Omega\!=\!\Omega_s\!+\!\delta \Omega,  ~\theta\!=\!\theta_s\!+\!\delta \theta$. 
Since $\dot \theta_s\!=\!0$, we have $\dot \theta\!=\!\delta \dot \theta$, and 
$\frac{dE^{(1)}}{d\theta}$ is  expanded as
\begin{equation}
\frac{dE^{(1)}}{d\theta}=\frac{d^2E^{(1)}}{d\theta^2}\biggr|_{\theta=\theta_s} \delta\theta 
+{\cal O}\Bigl( (\delta\theta)^2\Bigr)~,
\end{equation}
where we have used the fact $\frac{dE^{(1)}}{d\theta}\bigr|_{\theta\!=\!\theta_s}\!=\!0$. 

Meanwhile $v_{{\rm rot}PY}$ is shown to be expressed as 
\begin{equation}
v_{{\rm rot}PY}=V(n_s,\Omega_s,\theta_s)\delta\theta~. \label{vrotPYdelta}
\end{equation}
Actually we have already obtained the expressions (\ref{vPYthetazero}) and (\ref{vPYthetapi})
for $v_{{\rm rot}PY}(\approx v_{PY})$ near the steady states at $\theta\!=\!0$ and $\theta\!=\!\pi$, 
respectively. Also near the intermediate steady states,  $\delta n$ and $\delta\Omega$  are 
expressed in terms of $\delta\theta$ as (\ref{deltaNs}) and (\ref{deltaOmegas}), respectively, 
and thus we obtain (\ref{vrotPYdelta}). Now using Eqs.(\ref{EnergyEqLeading})-(\ref{vrotPYdelta}) 
we find 
\begin{equation}
\delta\dot\theta=-\mu Mg \frac{V(n_s,\Omega_s,\theta_s)^2}{|\vv_P(\Lambda)|}
\frac{1}{\frac{d^2E^{(1)}}{d\theta^2}\bigr|_{\theta=\theta_s} }\delta\theta~, 
\end{equation}
which means that we can identify $H$ in (\ref{EqForStability}) as
\begin{equation}
H\Bigl(n_s, \Omega_s, \theta_s, \frac{A}{C}, \frac{a}{R}\Bigr)=\frac{\rm a\ negative\ 
constant}{\frac{d^2E^{(1)}}{d\theta^2}\bigr|_{\theta=\theta_s}}~.
\end{equation}
Hence we conclude that the following assertions are equivalent:
{\it a steady state is stable (unstable) }$\Longleftrightarrow$
{\it $H$ is negative (positive) }$\Longleftrightarrow$
{\it $\frac{d^2E^{(1)}}{d\theta^2}\bigr|_{\theta=\theta_s}$ is positive (negative)}. 

In fact, ES showed that if the quantity (ES-(39)) with the upper sign is positive, then 
$\frac{d^2E^{(1)}}{d\theta^2}\bigr|_{\theta=\theta_s}$ is positive at $\theta_s\!=\!0$ 
and the non-inverted rotating motion is Liapunov stable. 
On the other hand, starting from the equations of motion we derived $H$ and obtained 
the condition (\ref{zeroStability2}) for the stability of the rotating motion 
at $\theta_s\!=\!0$.  It is easily seen that 
the statement that the quantity (ES-(39)) with the upper sign is positive is equivalent to 
the inequality given in (\ref{zeroStability2}), once we know that  Jellett's constant 
at $\theta_s\!=\!0$ is given by $J\!=\!Cn(\theta_s\!=\!0)(R\!-\!a)$.
Similarly, if the quantity (ES-(39)) with the lower sign is positive, then 
$\frac{d^2E^{(1)}}{d\theta^2}\bigr|_{\theta=\theta_s}$ is positive at $\theta_s\!=\!\pi$ 
and the completely inverted rotating motion is Liapunov stable. 
The condition that the quantity (ES-(39)) with the lower sign is positive is 
equivalent to the inequality given in (\ref{piStability2}). Note, this time, $J\!=\!Cn(\theta_s\!=\!\pi)(-R\!-\!a)$.

As for the intermediate steady state $(-1\!<\! \cos\theta_s\!<\!1)$, ES stated that if the steady state exists and 
the quantity (ES-(40))  is negative, then  $\frac{d^2E^{(1)}}{d\theta^2}\bigr|_{\theta=\theta_s}$ is 
positive and   the state is Liapunov stable. In Sec.4.2.3 we have shown  that the stability of the 
intermediate steady state is determined by the sign of $D(x_s)$ given in (\ref{Dxs}). 
Now it is interesting to note that $D(x_s)$ is related to (ES-(40)) as follows:
\begin{equation}
D(x_s)=-~ \frac{A^2+3[(A-C)x_s+C\frac{a}{R}]^2}{AC^2}\times ({\rm ES}.(40))~.
\end{equation}
Hence the condition that the quantity (ES-(40))  is negative is equivalent to $D(x_s)\!>\!0$. 
We have seen in Sec.4.2.3 that there exists an intermediate steady state for the tippe top of Group II and 
also of Group III.  (We have not considered a possible intermediate steady state for Group I, 
since such a state, even if it exists, cannot be reached from the initial spinning position 
near $\theta=0$.)  For these steady states, we have shown,  
in Appendix B, that   $D(x_s)$ is positive and,  therefore,  the states are stable.

\newpage
\section{Modified Maxwell-Bloch equations and stability criteria~\cite{BMR}}

Recently Bou-Rabee, Marsden and Romero [BMR] treated tippe top inversion as a dissipation-induced 
instability. 
They showed that the modified Maxwell-Bloch (mMB) equations are a normal 
form for tippe top inversion  and, using the mMB equations and an
energy-momentum argument,  they
gave criteria  for the stability on the non-inverted and inverted states of the tippe
top~\cite{BMR}.  
Although we have not explored the connections between the mMB equations and the 
first-order ODE (\ref{EqForStability}) for $\delta\theta$, 
we show in Appendix D that our results on the stability of the vertical spin states 
are consistent with the criteria provided by BMR. Actually,  rewritten  in terms of
dimensional parameters and classification criteria used 
in this paper,  the expressions of those criteria become more
transparent and they lead to the same  stability conditions as ours for the 
vertical spinning states. Besides, although BMR did not mentioned, 
the classification of tippe tops into three groups, Group I, II, and III, 
according to the  behaviors of spinning motion, is possible from the close examination of  those criteria.

BMR used the moments of inertia defined as the ones about the principal axes {\it attached to 
the center of sphere instead of the center of mass}.  The correspondence between the
parameters used by BMR and ones in this paper are as follows:
\begin{eqnarray}
|e^{\star}|&=&\frac{a}{R}~, \qquad \frac{1-\mu {e^{\star}}^2}{\sigma}=\frac{A}{C}~,  \label{BMRvsUSW}\\
\gamma_Q \Omega_{\rm BMR}&=&-\frac{J}{RC}~,  \qquad 
\frac{\mu |e^{\star}|Fr^{-1}}{\sigma}\Omega_{\rm BMR}^2=\frac{Mga}{C}~, \nn
\end{eqnarray}
where $\Omega_{\rm BMR}$ is the spin rate of the initially standing equilibrium 
solution (we added a subscript BMR to distinguish from our $\Omega$), 
and the dimensionless BMR's ``Jellett" constant, $\gamma_Q$, is restricted to have a certain value, i.e., 
$\gamma_Q\!=\!-(1+e^{\star})$.  Also BMR expressed the vector
from the center of sphere to the center of mass 
$\overrightarrow{SO}$ (in the BMR notation $\overrightarrow{OC}$) as $\overrightarrow{SO}\!=\!Re^{\star} \kv$, 
where $\kv$ is a unit vector along the symmetry axis. Using the tippe top modified 
Maxwell-Bloch equations, BMR obtained the stability criteria for the non-inverted state 
which are given by the three inequalities in (BMR-(5.3))\footnote{From now on, we write the equation ($\star\star$) 
of Ref.\cite{BMR} as BMR-($\star\star$). 
}.

They took $\kv\!=\!\ev_Z$ (upward)  in (BMR-(5.3)). Since the non-inverted state has 
the center of mass below the center of sphere, we have $e^{\star}\!=\!-|e^{\star}|\!=\!-\frac{a}{R}$, 
and thus $\Omega_{\rm BMR}\!=\!n(\theta\!=\!0)$. The first inequality of  (BMR-(5.3)) is 
rewritten as $\frac{A}{C}>0$, which is always satisfied. Apart from some irrelevant positive 
constants, the second and third inequalities are expressed,  respectively, as
\begin{eqnarray}
&& \frac{Mga}{[n(\theta=0)]^2C}\Bigl(1-\frac{a}{R}   \Bigr)\frac{A}{C}+\Bigl(1-\frac{a}{R}   \Bigr)^5
\frac{\nu^2}{\sigma^2}-\frac{A}{C}+\Bigl(1-\frac{a}{R}   \Bigr)>0~, \label{2ndIneq} \\
&&-\Bigl\{ \frac{A}{C}- \Bigl(1-\frac{a}{R}   \Bigr) -\frac{Mga}{[n(\theta=0)]^2C} 
\Bigl(1-\frac{a}{R}   \Bigr)^2\Bigr\}>0~. \label{3rdIneq}
\end{eqnarray} 
From these inequalities, we find: 
\begin{itemize}
\item [(ai)] In the case $\frac{A}{C}<(1-\frac{a}{R}   )$, i.e., for the tippe top of Group I, the above 
inequalities  are always satisfied. In other words, the non-inverted states ($\theta\!=\!0$) of Group I 
are always stable.
\item [(aii)] In the case $\frac{A}{C}>(1-\frac{a}{R}   )$, i.e., for the tippe tops of Group II or III, the 
inequality  (\ref{3rdIneq}) is satisfied  if 
\begin{equation}
[n(\theta=0)]^2<\frac{Mga}{C\{ \frac{A}{C} -(1-\frac{a}{R})\}}\Bigl(1- \frac{a}{R} \Bigr)^2, \label{D4}
\end{equation}
which is the same requirement given in (\ref{zeroStability3}) for the stability of 
the tippe top of Group II or III.  Note that the inequality  (\ref{2ndIneq}) is automatically
satisfied when both $\frac{A}{C}>(1-\frac{a}{R}   )$ and inequality  (\ref{3rdIneq}) hold. 
\end{itemize}
Thus, the BMR criteria (BMR-(5.3)) lead to the same result as ours on the 
 stability of the vertical spin state at $\theta=0$. 


The inequalities (BMR-(5.3)), which were derived as the stability criteria for the non-inverted 
state, can also be used for the  stability criteria for the inverted state, but with some replacements. 
Since $\kv\!=\!\ev_Z$ (upward), the inverted state has the  center of mass above the center of sphere. 
Thus we have $e^{\star}\!=\!\frac{a}{R}$ and $\Omega_{\rm BMR}\!=\!-n(\theta\!=\!\pi)$. 
Changing variables in inequalities (\ref{2ndIneq}) and (\ref{3rdIneq}) as 
$a\rightarrow -a$, $\frac{a}{R}\rightarrow
-\frac{a}{R}$, and 
$[n(\theta\!=\!0)]^2\rightarrow [n(\theta\!=\!\pi)]^2$, we obtain 
\begin{eqnarray}
&& -\frac{Mga}{[n(\theta=\pi)]^2C}\Bigl(1+\frac{a}{R}   \Bigr)\frac{A}{C}+\Bigl(1+\frac{a}{R}   \Bigr)^5
\frac{\nu^2}{\sigma^2}-\frac{A}{C}+\Bigl(1+\frac{a}{R}   \Bigr)>0~, \label{2ndIneqPi} \\
&&-\Bigl\{ \frac{A}{C}- \Bigl(1+\frac{a}{R}   \Bigr) +\frac{Mga}{[n(\theta=\pi)]^2C} 
\Bigl(1+\frac{a}{R}   \Bigr)^2\Bigr\}>0~, \label{3rdIneqPi}
\end{eqnarray} 
for the stability for the inverted state.
From the above two inequalities, we see: 
\begin{itemize}
\item [(bi)] In the case $\frac{A}{C}>(1+\frac{a}{R}   )$, i.e., for the tippe top of Group III, the 
inequality (\ref{3rdIneqPi})  is never  satisfied. 
Therefore, the inverted states ($\theta\!=\!\pi$) of Group III 
are always unstable.
\item [(bii)]  In the case $\frac{A}{C}<(1+\frac{a}{R}   )$, i.e., for the tippe top of Group I or II, the 
inequality  (\ref{3rdIneqPi}) is satisfied  if 
\begin{equation}
[n(\theta=\pi)]^2>\frac{Mga}{C\{ (1+\frac{a}{R}) -\frac{A}{C}\}}\Bigl(1+ \frac{a}{R} \Bigr)^2,
\label{D7}
\end{equation}
which is the same requirement given in (\ref{piStability3}) for the stability of 
the tippe top of Group I or II at $\theta\!=\!\pi$.  The inequality  (\ref{2ndIneqPi}) is automatically
satisfied when both $\frac{A}{C}<(1+\frac{a}{R}   )$ and inequality  (\ref{3rdIneqPi}) hold.
\end{itemize}
Thus, the BMR criteria (BMR-(5.3)) also lead to the same result as ours on the 
 stability of the vertical spin state at $\theta\!=\!\pi$. 

Actually, BMR  derived also the stability criteria for the inverted state, taking $\kv\!=\!-\ev_Z$, 
which are given by the three inequalities in (BMR-(5.4))\footnote{The second inequality
should read as
$\sigma(1\!+\! e^{\star})^2[\sigma(1\!-\! {e^\star})\!-\! (1\!-\! \mu {e^{\star
}}^2)]\!+\! \nu^2(1\!-\! e^\star)^7\!+\! (1\!-\! e^{\star})^3\mu e^{\star}Fr^{-1} (1\!-\!\mu
{e^{\star }}^2)>0$.   The error is traced back
to the missing factor of $(\gamma_z^0 n^0)$ in the expression  of $F$ in BMR-(4.2).}. Of course, we can
use them to obtain the stability conditions  for the inverted state. 
Taking now $\Omega_{\rm BMR}\!=\!-\frac{1\!-\! e^\star}{1\!+\! e^\star} n(\theta\!=\!\pi)$ and
$e^\star\!=\!-\frac{a}{R}$ in the second and third inequality in (BMR-(5.4)), we reach the same
conclusions, (bi) and (bii).

BMR discussed in Ref.~\cite{BMR} about the heteroclinic connection between the non-inverted and 
inverted states of the tippe top. 
%
%
They used an energy-momentum argument to determine the asymptotic states of the tippe top 
and obtained the explicit criteria for  the existence of a 
heteroclinic connection, which are given in Theorem 6.2  and the appendix of Ref.~\cite{BMR}. 
In terms of the classification criteria and conditions obtained in this paper, 
the statement in BMR on the existence of a
heteroclinic connection can be restated  as follows: 
(i) A tippe top must belong to Group II in order to have a heteroclinic connection.
(ii) Further more, the initial spin $n(\theta\!=\!0)$  should be larger than $n_1$ (Eq.(\ref{zeroStability4}))
and  $n_4$ 
(Eq.(\ref{critical1}))  so that 
a tippe top becomes unstable  at $\theta\!=\!0$ and  reaches  the inverted position. 
The requirements $n(\theta\!=\!0)\!>\! n_1$ and $n(\theta\!=\!0)\!>\! n_4$, respectively, 
correspond to the  criteria $r_0\!>\!0$ and $r_4\!>\!0$ in Theorem 6.2 in BMR.

\newpage

\end{document}